\documentclass[12pt]{article}

\usepackage{amsthm,amsfonts,amsmath,amssymb,amscd}
\usepackage{bm}
\usepackage{mathabx}
\usepackage{tabularx}
\usepackage{amsmath}
\usepackage[english]{babel}
\usepackage[protrusion=true,expansion=true]{microtype}
\usepackage[titletoc,toc,title]{appendix}	
\usepackage{graphicx}
\usepackage{caption}
\usepackage{multicol}
\usepackage{subcaption}
\usepackage[colorinlistoftodos]{todonotes}
\usepackage{mathtools}
\usepackage{changepage}
\usepackage{placeins}
\usepackage{float}
\usepackage[nottoc]{tocbibind}
\usepackage{varwidth}
\usepackage{textcomp}
\newcommand*{\No}{\textnumero}

\usepackage{authblk}

\begin{document}
	\title{Random walks and market efficiency in Chinese and Indian equity markets}
	\author[1]{Oleg Malafeyev\thanks{malafeyevoa@mail.ru}}
	\author[2]{Achal Awasthi\thanks{aa777@snu.edu.in}}
	\author[2]{Kaustubh S.Kambekar\thanks{kk554@snu.edu.in}}
	\affil[1]{Saint-Petersburg State University,  Russia}
	\affil[2]{Shiv Nadar University, India}
	
\date{}
\maketitle

\begin{abstract}
Hypothesis of Market Efficiency is an important concept for the investors across the globe holding diversified portfolios. With the world economy getting more integrated day by day, more people are investing in global emerging markets. This means that it is pertinent to understand the efficiency of these markets. This paper tests for market efficiency by studying the impact of global financial crisis of 2008 and the recent Chinese crisis of 2015 on stock market efficiency in emerging stock markets of China and India. The data for last 20 years was collected from both Bombay Stock Exchange (BSE200) and the Shanghai Stock Exchange Composite Index and divided into four sub-periods, i.e. before financial crisis period (period-I), during recession (period-II), after recession and before Chinese Crisis (period- III) and from the start of Chinese crisis till date (period- IV). Daily returns for the SSE and BSE were examined and tested for randomness using a combination of auto correlation tests, runs tests and unit root tests (Augmented Dickey-Fuller) for the entire sample period and the four sub-periods.

The evidence from all these tests supports that both the Indian and Chinese stock markets do not exhibit weak form of market efficiency. They do not follow random walk overall and in the first three periods (1996 till the 2015) implying that recession did not impact the markets to a great extent, although the efficiency in percentage terms seems to be increasing after the global financial crisis of 2008. 
\end{abstract}

\section{Overview}
Previous research shows us that the strong functioning of stock markets has considerable effect on the growth of an economy, especially so in a developing one. Over the past few decades, studies have been conducted around the globe by many researchers on the subject of stock market efficiency. And the conflicting results have made it difficult to comment on the status of stock market of a particular country. So, we focus our attention on the stock market behaviour in developing countries which aren’t considered to be as stable as the developed ones. They are unlikely to be fully information-efficient, partly due to institutional barriers restricting information flows to the market and partly due to lack of experience of market participants to rapidly lock up new information into security prices. Therefore, it would be interesting to investigate this period of last 20 years studying both the Global Financial and the Chinese crisis and its effects on fastest emerging economies of India and China. 
\\
\\
Recession had crumpled economies worldwide but these two were relatively unaffected and hence are of particular interest. The the current fastest growing economies BRICS (Brazil, Russia, India, China, South Africa) were affected primarily through four channels of trade, finance, commodity, and confidence. The slump in export demand and firmer trade credit caused a slowdown in aggregate demand. The global financial crisis inflicted significant loss in output in all these countries. However, the real GDP growth in India and China remained impressive even though they witnessed some moderation due to weakening global demand. The crisis also exposed the structural weakness of the global financial and real sectors. The BRICS were able to recover quickly with the support of domestic demand. The reversal of capital flows led to equity market losses and currency depreciations, resulting in lower external credit flows. The banking sectors of the BRICS economies performed relatively well. 
\\
\\
So, we will try to answer the question of market efficiency in these countries through this study. The remainder of this study is organized as follows. Section 2 provides introduction to efficient market hypothesis, random walk theory, financial crisis and theirs effects on India and China. Section 3 reviews the previous studies on weak-form efficiency in developed and developing economies. Section 4 discussed the objectives, scope and rationale of the study undertaken. Section 5 describes the data, hypotheses and methodology of the empirical research. Section 6 presents the empirical results and section 7 provides limitations of this study and draws conclusion.

\section{Introduction}

The efficient market hypothesis (EMH), popularly known as the Random Walk Theory, is concerned with the informational efficiency of the capital markets. In the literature of capital markets, the term market efficiency is used to explain the relationship between information and share prices. It was first introduced and defined by Eugene Fama in 1970, where he had defined market efficiency as the efficiency in stock markets when the security prices in that market adjust rapidly to the introduction of new information. Hence, in any efficient market, current prices of securities should reflect all the information useful for price prediction of securities in the stock market and there is no way to earn excess profits (more than the market) by using this information. This depends upon the extent to which the information is absorbed, the time taken for absorption and the type of information absorbed. The price of the security reflects the present value of its expected future cash flows, which incorporates many factors such as volatility, liquidity, and risk of bankruptcy. However, while the prices are rationally based, changes in prices are expected to be random and unpredictable, because of the fact that new information, is unpredictable by its very nature. Therefore, stock prices are said to follow a random walk.
\\
\\
Random walk theory implies that statistically stock price fluctuations have the same distribution, are independent of each other, and may be described by a random process so the past movement or trend of a stock price or market cannot be used to predict its future movement. Tossing of a coin or selection of a sequence of numbers from a random number table are few examples of a random process. This posits hat the current market price of a given stock is independent of and unrelated to previous market price patterns. A follower of the random walk theory believes it's impossible to outperform the market without assuming additional risk. This theory implies that a series of stock price changes has no memory- that one cannot predict future market prices on the basis of past history of price behaviour. Critics of the theory, however, contend that stocks do maintain price trends over time - in other words, that it is possible to outperform the market by carefully selecting entry and exit points for equity investments.
\\
\\
A stochastic variable X is said to follow a random walk if:

\begin{equation}
X(t+1) = \delta + X(t) + \epsilon_{t+1}
\end{equation}
with $\delta$ being a drift parameter and the forecast error $\epsilon_{t+1}$ being identically and independently distributed. A random walk without drift has $\delta$ = 0.
\\
\\
Fama (1970) suggested that the efficient market hypothesis can be divided into three categories based on the type of information that is fully reflected in the security prices:

\begin{itemize}
	
	\item The weak form of EMH assumes that current stock prices fully reflect all currently available security market information. It states that past price and volume data have no relationship with the future movement of security prices and that excess returns cannot be achieved using fundamental analysis.
	
	\item The semi-strong form of EMH assumes that current stock prices adjust rapidly to the release of all new public information. It states that security prices have all available market and non-market public information factored in and that excess returns cannot be achieved using fundamental analysis.

	\item The strong form of EMH assumes that current stock prices fully reflect all public and private information. It states that market, non-market and inside information is all factored into security prices and that no one has monopolistic access to relevant information. It assumes a perfect market and concludes that excess returns are impossible to achieve consistently.
	
\end{itemize}

\noindent

The theoretical foundations for the efficient market hypothesis rest on the following three arguments:
\begin{itemize}
	\item Investor rationality: Investors are assumed to be rational, which means that they correctly update their beliefs when new information is available.
	\item Arbitrage: Even if all the investors are assumed not to be rational, some rational investors use arbitrage to remove pricing errors, so the average investor would not matter; the marginal investor is the one setting prices. 
	\item Collective rationality: The random errors of investors cancel out in the market. Some investors are not rational, they trade randomly and, consequently, their trades cancel each other without affecting the prices.
\end{itemize}
Since our analysis of market efficiency revolves around the two recent financial crises, we need to understand its effects as well. A financial crisis is a disruption to financial markets in which adverse selection and moral hazard problems become much worse, so that financial markets are unable to efficiently channel funds to those who have the most productive investment opportunities. As a result, a financial crisis can drive the economy away from an equilibrium with high output in which financial markets perform well to one in which output declines sharply (Mishkin, 1992). The end of 2007 and beginning of 2008 observed that the onset of global financial crisis had bought disorder to the financial markets around the world and it is the first crisis in consideration for our study. The instability in the global stock market scenario began with a shortfall of liquid assets in US banking system and the continual fall in stock prices on information that Lehman Brothers, Merill Lynch and many other investment banks and companies are collapsing. The stock markets around the globe suffered huge losses and Indian stock market was no exception. The SENSEX which had reached historically high levels in the beginning of 2008, turned down to its level about three years back and the S$\&$P CNX NIFTY also followed the similar trend. Economic growth decelerated in 2008-09 to 6.7 percent. This represented a decline of 2.1 percent from the average growth rate of 8.8 percent in the previous five years.
\\
\\
China was not one of the countries hardest hit by the crisis, neither was it as insulated as many had assumed. This can be seen from the fact that China continued to have one of the
highest rates of economic growth across the globe, recording 9.6$\%$ in 2008 and 9.2$\%$ in 2009. Although while most countries would be delighted to have such growth rates, the point to be considered is that these rates reflected a substantial drop from the 14.2$\%$ growth in 2007. In terms of short term impact on China, the most visible damage was inflicted on its export-oriented light industry in southern China. Thousands of companies went bust, tens of thousands of workers have been laid-off and official statistics revealed that 10 million migrant workers had returned back to their home provinces. In the financial sector the stock market crash that started in late 2007 had wiped out more than two thirds of market value although this dramatic collapse was not without any home-made reasons. The Chinese banks for all their profitability witnessed the sudden pull-out of many of their Western partners which (Bank of America, UBS, RBS) sold their minority stakes in order to retrieve capital. Another massive blow was to the China's fledgling sovereign wealth fund, China Investment Corporation. 
\\
\\
The second crisis in consideration for our study is the Chinese stock market crash which began with the popping of the stock market bubble on 12 June 2015. A third of the value of A-shares on the Shanghai Stock Exchange was lost within one month of the event since mid-June. By 8–9 July 2015, the Shanghai stock market had fallen 30 percent over three weeks as 1,400 companies, or more than half listed, filed for a trading halt in an attempt to prevent further losses.
This crisis was inevitable because over major part of 2014-15, investors kept investing more and more into Chinese stocks, encouraged by falling borrowing costs as the central bank loosened monetary policy even though economic growth and company profits were weak with retail investors being the one leading this.

\section{Literature Review}

Over recent decades, there has been a large body of empirical research concerning the validity of the random walk hypothesis with respect to stock markets in both developed and developing countries. Empirical research on testing the random walk hypothesis has produced mixed results. Various researchers have given their observations and views regarding different stock markets worldwide. The results vary and are often in contradiction with one another giving us two different schools of thoughts, the first supporting the presence of weak form market efficiency in some stock markets and the other denying the same time and claiming no evidence of random walk in the same or some other stock market worldwide.

\subsection{Studies Supporting Weak Form Market Efficiency}
Fama (1965) found that the serial correlation coefficients for a sample of 30 Dow Jones Industrial stocks, even though statistically significant, were too small to cover transaction costs of trading. Chen (2008) uses three variance ratio tests: Lo-MacKinlay’s conventional variance ratio test, Chow-Denning’s simple multiple variance ratio test, and Wright’s non-parametric ranks and signs based variance ratio tests to examine the random walk hypothesis (RWH) of the Euro/U.S. Dollar exchange rate market using the data from January 1999 to July 2008. The results show that the RWH cannot be rejected for all of three variance ratio tests’ and hence find the Euro/U.S. Dollar exchange rate market as weak-form efficient.
\\
\\
Chan, Gup, and Pan (1992) analyzed the weak form hypothesis in Hong Kong, South Korea, Singapore, Taiwan, Japan, and the United States. Their findings indicate that stock prices in these major Asian markets and the United States are efficient in the weak form. Worthington and Helen (2004) tests for random walks and weak-form market efficiency in European equity markets. Daily returns for sixteen developed markets and four emerging markets across Europe are examined for random walks using a combination of serial correlation coefficient and runs tests, Augmented Dickey-Fuller (ADF), Phillips-Perron (PP) and Kwiatkowski, Phillips, Schmidt and Shin (KPSS) unit root tests and multiple variance ratio (MVR) tests. The results, which were in broad agreement across the approaches employed indicated that of the emerging markets only Hungary was characterized by a random walk and hence is weak-form efficient, while in the developed markets only Germany, Ireland, Portugal, Sweden and the United Kingdom comply with the most stringent random walk criteria.
\\
\\
Mahmood et al. (2010) tried to examine the impact of recent financial crisis on the efficiency of Chinese stock market by dividing the stock price data from Shanghai and Shenzhen stock market for the period of six years, starting from January 2004 to December 2009, into two sub-periods, i.e. before crisis and during crisis period. The sample data was analyzed by applying Runs test, Variance Ratio test, Durbin-Watson test and Unit Root (ADF) test and it was concluded that the Chinese stock market was weak form efficient and global financial crisis has no significant impact on the efficiency of Chinese stock market.
\\
\\
Wu (1996) examine efficiency in both Chinese stock markets, on the early stage of development in Shanghai and Shenzhen stock exchanges. Using the serial correlation test on eight and twelve individual shares for the period from June 1992 to December 1993, he finds Chinese stock markets to be weak-form efficient.
There are number of studies wherein researchers tried to test the Indian stock market's behavior in terms of stock market efficiency. Vaidyanathan and Gali(1994) tested for the weak form efficiency checking for randomness using the runs test, serial correlation and filter rule tests based on the daily closing prices of ten shares actively traded on the Bombay Stock Exchange and found evidence from all the three tests supporting the weak form of Efficient Market Hypothesis.
\\
\\
Other researches includes the studies done by Bhaumik (1997), Rao and Shankaraiah (2003), Sharma and Mahendru (2009), who ran various econometric tests, found that the returns follow random walk and concluded that the BSE is weak form efficient. Pant and Bishnoi (2002) have also used autocorrelation function, unit root test and variance ratio to examine the random walk hypothesis for Indian stock market. On the basis of the test results, they concluded that Indian stock market follows random walk and thus is efficient in weak form. Mall, Pradhan, and Mishra (2011) use daily data from June 2000 to May 2011 and found that the Indian capital market is weak form efficient.

\subsection{Studies Not Supporting Weak Form Market Efficiency}
Smith and Ryoo (2003) tested the hypothesis that stock market price indices follow a random walk using the multiple variance ratio test for five European emerging markets, Greece, Hungary, Poland, Portugal and Turkey. In four of them, the random walk hypothesis is rejected because of autocorrelation in returns. Lo and MacKinlay (1988) strongly rejected
random walk model for a sample of 1216 weekly observations of firms in the NYSE-AMEX over the period 1962-1985. Frennberg and Hansson (1993) checked for the random walk hypothesis on monthly data for the Swedish stock market for 1919- 1990 using variance ratio test and autoregression of multiperiod returns. They found that the Swedish markets have not followed random walk in past 72 years. An interesting result was strong evidences of positive autocorrelated returns were found for short investment horizons whereas, for long investment horizons indication of negative autocorrelation was present. 
\\
\\
Anil K. Sharma and Neha Seth (2011) studied the impact of recent financial crisis on stock market efficiency in India. The data for last 10 years were collected from both Bombay Stock Exchange (BSE) and National Stock Exchange (NSE) in India. The data was divided into two sub-periods, i.e. before financial crisis period (period-I) and during financial crisis period (period-II) and it was concluded that Indian stock markets do not exhibit weak form of market efficiency and thus do not follow random walk in both period-I and period-II. It was found that the recent financial crisis did not impact the behavior of Indian stock markets to a great extent. Harper, Alan; Zhenhu Jin(2012) concluded that the Indian stock market was not weak form efficient from 1997-2011 using using autocorrelation, the Box-Ljung test statistics and the runs test. Gupta and Basu(2007) also founnd evidence suggesting the Indian Stock market does not follow random walk model and that there is an evidence of autocorrelation in both BSE and NSE markets rejecting the weak form efficiency hypothesis. 
\\
\\
Niblock and Sloan(2007) evaluate whether or not Chinese stock markets are weak-form efficient, based on analysis of daily data of the Shanghai “A”, Shanghai “B”, Shenzhen “A”, Shenzhen “B”, Hang Seng, and Dow Jones Industrial Average indices from 2002 to 2005. Tests of the random walk hypothesis reveal return predictabilities for the Chinese share indices together with some evidence of increased predictability in the most recent period. The results supported the assertion that despite continual financial liberalization and unparalleled growth, China’s stock markets were still not weak-form efficient.

\section{Objective and Rationale}

The aim of this paper is to understand the nature of the Indian and Chinese stock market by testing the randomness in the daily market returns and to study the impact of the two recent financial crisis. Randomness in stock returns is possible only if market is efficient in weak form. Hence, we check whether these markets follow random walk to test for weak form market efficiency. The studies on market efficiency of Indian and Chinese markets especially during the crises are very few. Through this study we plan to test whether these markets are weak form efficient or not and add to the existing literature. 
\\
\\	
EMH uses past actual prices or returns for the tests. Sets of share price changes are tested for serial independence. As discussed earlier there are three forms of market efficiency, but working on all the three forms is not possible because of unavailability of the data and usefulness of the results in acheiving the objectives of the study. The results for testing semi-strong form do not measure the randomness in the market returns, which would only be possible through testing the weak form of market efficiency. On the other hand, for strong form of market efficiency, the tests are not possible because of unavailability of data as it considers private or insider information which is not easy to access. Our objective here is to test the randomness in the market returns before, during and after the financial crisis, whether these had any effect on the market returns which is possible by testing market efficiency in weak form.
\\
\\
Valuation of securities is an important function of financial markets which leads to the formation of investment strategies for the traders and investors trading in these markets. The valuation of securities is required to identify the behaviour of the markets, which is again possible by knowing the status of efficiency in the stock market. This information is crucial as it could lead to possible arbitrage opportunities. Arbitrage is the "simultaneous purchase and sale of the same, or essentially similar, security in two different markets for advantageously different prices” (Sharpe and Alexander (1990)). It requires no capital and is riskless profit. When an arbitrageur buys a cheaper security and sells a more expensive one, his net future cash flows are zero for sure, and he gets his profits up front. Arbitrage plays a critical role in the analysis of securities markets, because its effect is to bring prices to fundamental values and to keep markets efficient. In the context of the stock market, traders often try to exploit arbitrage opportunities. For example, a trader may buy a stock on a foreign exchange where the price has not yet adjusted for the constantly fluctuating exchange rate. The price of the stock on the foreign exchange is therefore undervalued compared to the price on the local exchange, and the trader makes a profit from this difference.
\\
\\
If the markets are efficient in weak form, buying the undervalued securities and selling them into the markets, where securities are fairly valued or overvalued, is not possible. This makes it difficult for the investors to earn abnormal profits on trading in such markets. Hence, weak form market efficiency leads to the decisions related to buying and selling of the undervalued or overvalued securities on right time.

\section{Data}
Data of daily closing prices of SSE composite index and BSE200 for the period Jan 1, 1996 to April 8, 2016 is considered for the study. The data for SSE was retrieved from Yahoo! Finance whereas the one for BSE200 was taken from its official website, bseindia.com.
\\
For this purpose, we have used the daily closing prices BSE200 from BSE and the adjusted closing prices for the SSE Composite Index. These markets, BSE and SSE are considered because of their popularity around the world so as to represent these markets. 
\\
\\
S$\&$P BSE 200 Index is a free float weighted index of 200 companies selected from Specified and Non-Specified lists of BSE India Exchange, selected based on their market capitalization. It started as a cap-weighted index with a base value of 100, and base year 1989-90. Effective from 8/16/05, it was changed to a free float index. Though S$\&$P BSE SENSEX was serving the purpose of quantifying the price movements as also reflecting the sensitivity of the market in an effective manner, the rapid growth of the market necessitated compilation of a new broad-based index series reflecting the market trends in a more effective manner and providing a better representation of the increased equity stocks, market capitalization as also to the new industry groups. As such, BSE launched on 27th May 1994, two new index series S$\&$P BSE 200 and S$\&$P Dollex 200. The equity shares of 200 selected companies from the specified and non-specified lists of BSE were considered for inclusion in the sample for `S$\&$P BSE 200'. The selection of companies was primarily done on the basis of current market capitalization of the listed scrips. Moreover, the market activity of the companies as reflected by the volumes of turnover and certain fundamental factors were considered for the final selection of the 200 companies. 
\\
\\
We have selected BSE 200 because of the diversification that it would offer in terms of the companies that are represented on this index reducing sampling error due to the presence of a large number of companies. This would result in heterogeneity as the Sensex Index comprises of only the top performing companies which wouldn’t be representative of the market.
\\
\\
The Chinese stock market has been one of the fastest growing emerging capital markets, and is now the second largest in Asia, only behind Japan. The Shanghai Stock Exchange Composite Index is a capitalization-weighted index. The index tracks the daily price performance of all A-shares and B-shares listed on the Shanghai Stock Exchange. The index was developed on December 19, 1990 with a base value of 100. The first day of reporting was July 15, 1991. 
\\
\\
\textbf{A-shares} are shares of the Renminbi currency that are purchased and traded on the Shanghai and Shenzhen stock exchanges.
\\
This is contrast to Renminbi \textbf{B-shares} which are owned by foreigners who cannot purchase A-shares due to Chinese government restrictions. 
\\
\textbf{B shares} (officially Domestically Listed Foreign Investment Shares) on the Shanghai and Shenzhen stock exchanges refers to those that are traded in foreign currencies.
\\
\\
The composite figure can be calculated by using the formula:
$$\mathrm{Current\; Index}=\frac{\mathrm{Market\; Cap\; of\; Composite\; Numbers}}{\mathrm{Base\; Period}} \times \mathrm{Base\; Value}$$
\\
\noindent

The B-share stocks are generally denominated in US dollars for calculation purposes. For calculation of other indices, B share stock prices are converted to RMB at the applicable exchange rate (the middle price of US dollar on the last trading day of each week) at China Foreign Exchange Trading Center and then published by the exchange.
\\
\\
According to the US National Bureau of Economic Research (the official arbiter of US recessions) the US recession began in December 2007 and ended in June 2009, and thus extended over 19 months.
In order to see the impact of recent financial crisis and have time varying results, the total data is divided into four equal sub-periods, i.e. before financial crisis period (period-I, Jan1996- Nov2007), during recession (period-II, Dec2007- Jun2009), after recession and before Chinese Crisis (period- III, Jul2007- May2015) and from the start of Chinese crisis till date(period- IV, June 2015- Apr 2016). We assume the sample period is sufficient to evaluate the information asymmetry especially after the huge Foreign Institutional Investors investments in stock markets, sub-prime crisis disorder and the recent financial crisis.
\\
\\

\section{Research Methodology}

This study applies a classical framework of testing market efficiency by determining whether or not the time series data from Chinese and Indian stock returns follow the random walk model using statistical tests including normality, autocorrelation, runs test, variance ratio and unit root tests for null hypothesis of a random walk are employed:
\\
\\
\textbf{Jarque-Bera (JB) Test} is used to measure the normality of the distribution. The results of skewness and kurtosis, which are also used to test the normality, are verified by the JB test. So, JB test is the test of joint hypothesis that skewness, S and kurtosis, K are zero and three respectively. The value of JB statistic is calculated by using the following equation:
$$\mathrm{JB} = n\cdot \left( \frac{s^2}{6} + \frac{(K-3)^2}{24} \right) $$
The conclusion can also be drawn on the basis of probability value. If the value of probability is more that 0.05 at 5\% level of significance, we accept the null hypothesis for normality and can conclude that the observed series follows normal distribution.
\\
\\
\textbf{Runs Test} is used to test randomness of the time series. The null hypothesis of the test is that the observed series is random variable. When the Z value is more than + 1.96 at 5 percent level of significance, the test rejects the null hypothesis. The test is non-parametric and is independent of the normality and constant variance of data.
\\
\\
\textbf{Autocorrelation Function (ACF)} is used to determine the correlation of series with itself. It measures the amount of linear dependence between observations in a time series that are separated by lag. The ACF shows the pattern of autocorrelation present in time series as well as the extent to which current values of the series are related to various lags of the past data. ACF shows whether the serial correlation is significantly different from zero. In an efficient market, existence of zero autocorrelation forms the null hypothesis.
\\
\\
\textbf{Unit Root Test} is used to test the stationarity of time series data and to find out whether a time series variable is non-stationary. The most popular unit root tests used to test the stationarity are the Augmented Dickey-Fuller (ADF) test and Phillips-Perron (PP) test. Both tests use the existence of a unit root as the null hypothesis. If the series is nonstationary and the first difference of the series is stationary, the series contains a unit root. 
\\
\\
Return of the indexes are used to conduct the run test, normality test and autocorrelation for the time series, Daily return has been calculated as follows:
$$r = \frac{P_t-P_{t-1}}{P_{t-1}}$$
It may also be noted that the price of the indexes are used to test the stationarity.
\\
\\
The main objective of this study is to examine whether the stock market follows a random walk or is weak form efficient. Accordingly, the hypothesis of the study is: 
\begin{itemize}
	\item $H_0$: The stock market follows a random walk is a weak-form efficient 
	\item $H_1$: The stock market does not follow the random walk
\end{itemize}

\section{Analysis and Findings}

\subsection{Descriptive Statistics}

The descriptive statistics below in table 1 and 2 describes the characteristics of the data for the entire time frame and all the four periods. The sample means, median, maximum, standard deviation, skewness, kurtosis, Jarque-Bera statistics and probability value are given here under.
\\
From the data, it can be inferred that BSE has higher values in terms of mean,  maximum price than the values of SSE composite Index but the max values for returns for SSE is greater after the global financial crisis. higher values of standard deviation over the periods explain that the SSE is more volatile market as compared to BSE. The volatility for both the markets had increased during the period of financial crisis. The values of skewness (skewness of the normal distribution or any perfectly symmetric distribution is zero) and kurtosis (kurtosis of the normal distribution is 3) reveal that both markets do not follow normal distribution, which is further verified by the value of Jarque-Bera statistic and probability value.

\FloatBarrier
\begin{table}[!htbp]
	\begin{adjustwidth}{0.0cm}{}
		\resizebox{\columnwidth}{!}{%
			\begin{tabular}{c|c|c|c|c|c} \hline
				& Daily & Daily (Period 1) & Daily (Period 2) & Daily (Period 3) & Daily (Period 4)\\ \hline
				Start & 01/01/1996	& 01/01/1996	& 12/03/2007	&01/07/2009	&01/06/2015 \\ \hline
				End	& 08/04/2016 &	30/11/2007	&30/06/2009	&29/05/2015	&08/04/2016\\ \hline
				Observations & 5153	& 3092 &	399	& 1465 & 197\\ \hline
				Mean returns & 4.7297e-04&8.4008e-04&-9.3764e-04&3.9179e-04&-0.0018\\ \hline
				Max. returns&0.0986&0.0986&0.0945&0.0479&0.0769\\ \hline
				Min. returns&-0.0991&-0.0991&-0.0773&-0.0770&-0.0849\\ \hline 
				Std. deviation&0.0171&0.0166&0.0250&0.0133&0.0274\\ \hline 
				Skewness&-0.1836&-0.0466&0.1667&0.4712&-0.5696\\ \hline
				Kurtosis&7.9450&8.8253&4.5031&6.0342&4.0450\\ \hline
				Jarque -Bera&5.2791e+03&4.3730e+03&1.4167e-04&616.1753&19.6151\\ \hline
				JB p-value& 0.0001*	&0.0001*&0.0001*&0.0001*&0.0001*\\ \hline
			\end{tabular}%
		}
		\caption{Descriptive statistics for the returns of the SSE composite stock index}\label{tbl:names}
	\end{adjustwidth}
\end{table}
\FloatBarrier
* means that the test rejects the null hypothesis the data comes from a normal distribution

\FloatBarrier
\begin{table}[!htbp]
	\begin{adjustwidth}{0.0cm}{}
		\resizebox{\columnwidth}{!}{%
			\begin{tabular}{c|c|c|c|c|c} \hline
				& Daily & Daily (Period 1) & Daily (Period 2) & Daily (Period 3) & Daily (Period 4)\\ \hline
				Start & 01/01/1996	& 01/01/1996	& 12/03/2007	&01/07/2009	&01/06/2015 \\ \hline
				End	& 08/04/2016 &	30/11/2007	&30/06/2009	&29/05/2015	&08/04/2016\\ \hline
				Observations & 5032&2965&383&1471&215\\ \hline
				Mean returns & 5.8424e-04&8.2236e-04&-4.9238e-04&5.2854e-04&-4.0426e-04\\ \hline
				Max. returns&0.1631&0.0794&0.1631&0.0373&0.0317\\ \hline
				Min. returns&-0.1187&-0.1187&-0.1073&-0.0569&-0.0651\\ \hline 
				Std. deviation&0.0157&0.0160&0.0271&0.0107&0.0109\\ \hline 
				Skewness&-0.1728&-0.3973&0.3402&-0.1855&-1.0274\\ \hline
				Kurtosis&9.1696&6.5587&6.8511&4.2371&8.4307\\ \hline
				Jarque -Bera&8.0042e+03&1.6421e+03&244.0618&102.1698&300.6266\\ \hline
				JB p-value& 0.0001*	&0.0001*&0.0001*&0.0001*&0.0001*\\ \hline
			\end{tabular}%
		}
		\caption{Descriptive statistics for the returns of the BSE 200 stock index}\label{tbl:names}
	\end{adjustwidth}
\end{table}
\FloatBarrier
* means that the test rejects the null hypothesis the data comes from a normal distribution

\begin{figure}[H]
	\centering
	\includegraphics[width=0.6\textwidth]{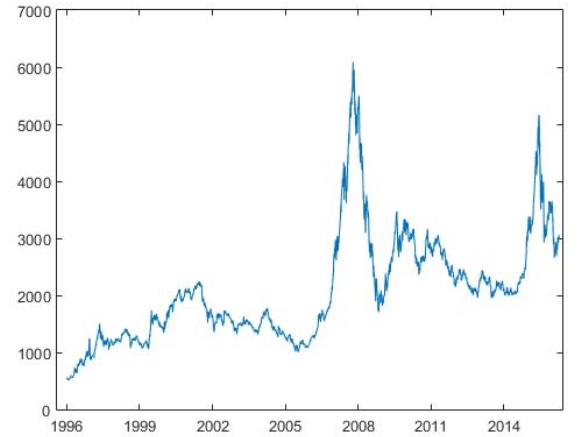}
	\caption{\label{fig:frog} SSE Composite Index evolution from 01/96 - 04/16}
\end{figure}

\begin{figure}[H]
	\centering
	\includegraphics[width=0.8\textwidth]{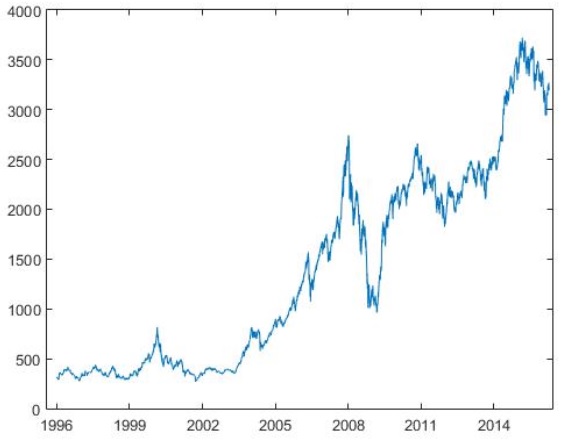}
	\caption{\label{fig:frog} BSE 200 index evolution from 01/96 - 04/16}
\end{figure}

\FloatBarrier
\begin{figure} [H]   
	\begin{minipage}[t]{0.45\textwidth}
		\includegraphics[width=\linewidth]{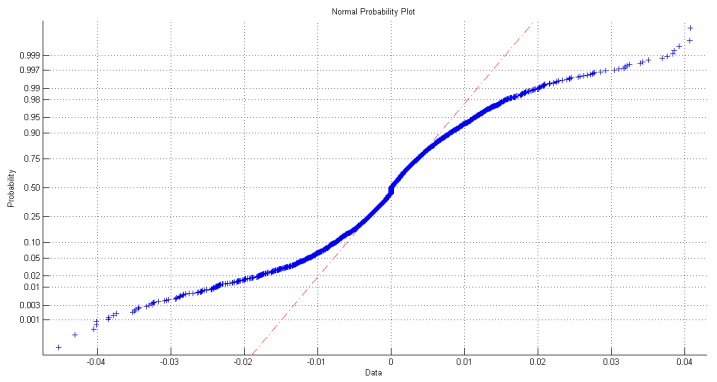}
		\caption{Daily Returns for SSE}
		\label{fig:immediate}
	\end{minipage}
	\hspace{\fill}
	\begin{minipage}[t]{0.45\textwidth}
		\includegraphics[width=\linewidth]{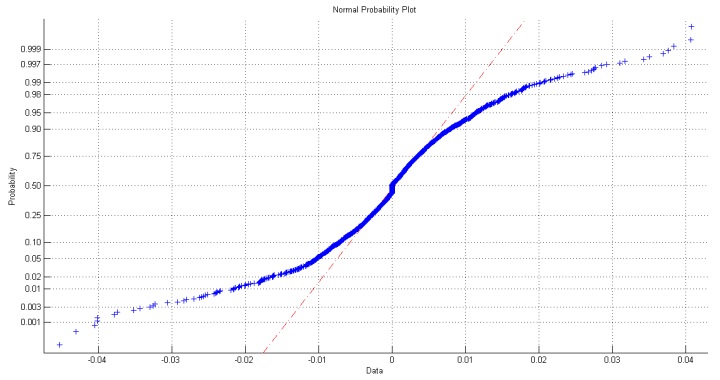}
		\caption{Daily Returns of SSE for period 1 (01/96 - 11/07) }
		\label{fig:proximal}
	\end{minipage}
	
	\vspace*{0.5cm} 
	\begin{minipage}[t]{0.45\textwidth}
		\includegraphics[width=\linewidth]{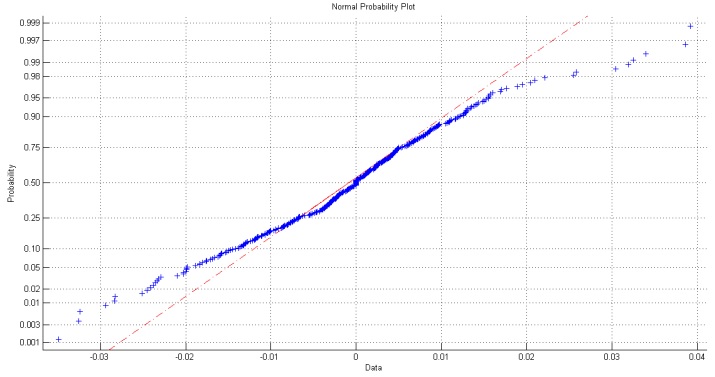}
		\caption{Daily Returns of SSE for period 2 (12/07 - 06/09) }
		\label{fig:distal}
	\end{minipage}
	\hspace{\fill}
	\begin{minipage}[t]{0.45\textwidth}
		\includegraphics[width=\linewidth]{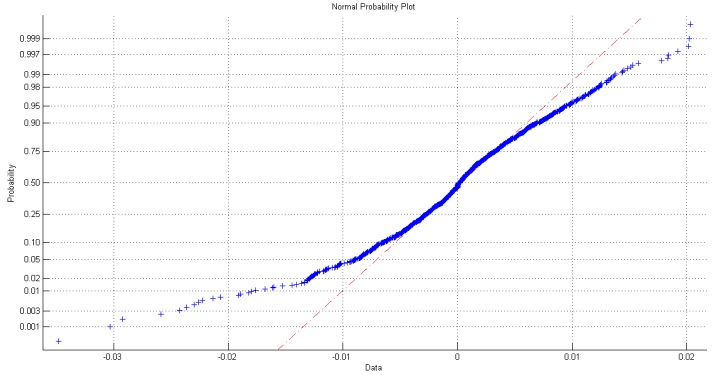}
		\caption{Daily Returns of SSE for period 3 (07/09 - 05/15) }
		\label{fig:combined}
	\end{minipage}
	\hspace{\fill}
	\begin{minipage}[t]{0.45\textwidth}
		\includegraphics[width=\linewidth]{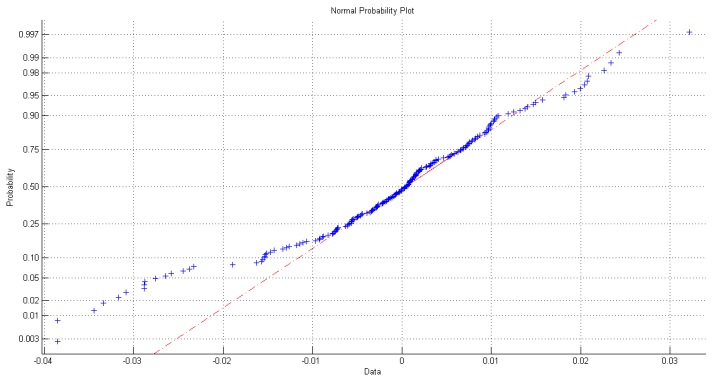}
		\caption{Daily Returns of SSE for period 4 (06/15 - 04/16)}
		\label{fig:combined}
	\end{minipage}
\end{figure}
\FloatBarrier

\FloatBarrier
\begin{figure} [H]   
	\begin{minipage}[t]{0.45\textwidth}
		\includegraphics[width=\linewidth]{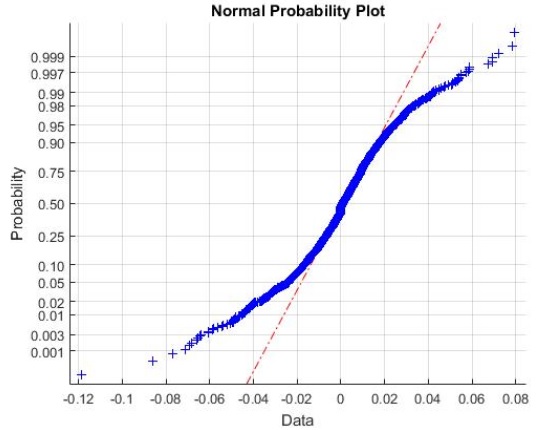}
		\caption{Daily Returns for BSE 200}
		\label{fig:immediate}
	\end{minipage}
	\hspace{\fill}
	\begin{minipage}[t]{0.45\textwidth}
		\includegraphics[width=\linewidth]{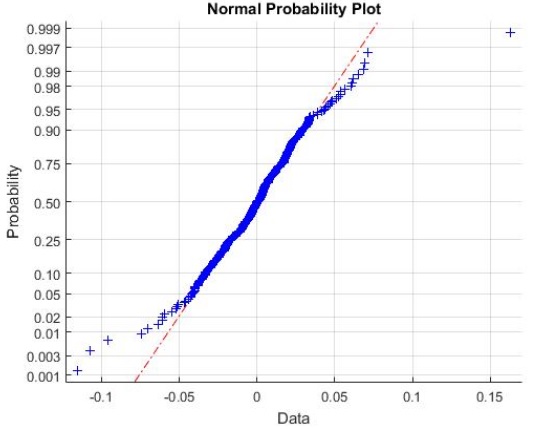}
		\caption{Daily Returns of BSE 200 for period 1 (01/96 - 11/07) }
		\label{fig:proximal}
	\end{minipage}
	
	\vspace*{0.5cm} 
	\begin{minipage}[t]{0.45\textwidth}
		\includegraphics[width=\linewidth]{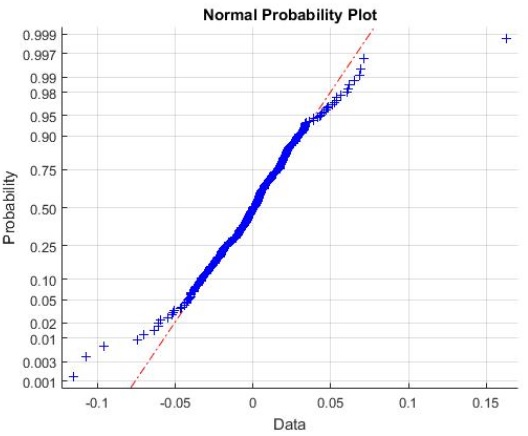}
		\caption{Daily Returns of BSE 200 for period 2 (12/07 - 06/09) }
		\label{fig:distal}
	\end{minipage}
	\hspace{\fill}
	\begin{minipage}[t]{0.45\textwidth}
		\includegraphics[width=\linewidth]{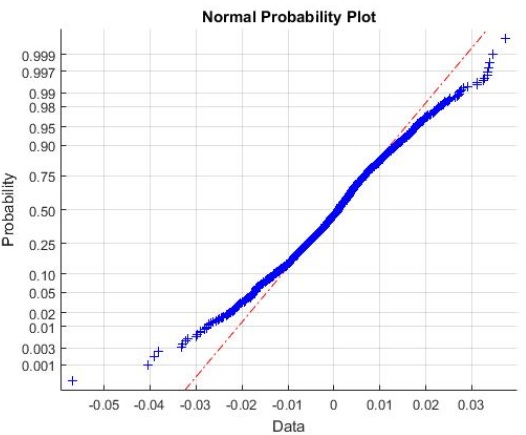}
		\caption{Daily Returns of BSE 200 for period 3 (07/09 - 05/15) }
		\label{fig:combined}
	\end{minipage}
	\hspace{\fill}
	\begin{minipage}[t]{0.45\textwidth}
		\includegraphics[width=\linewidth]{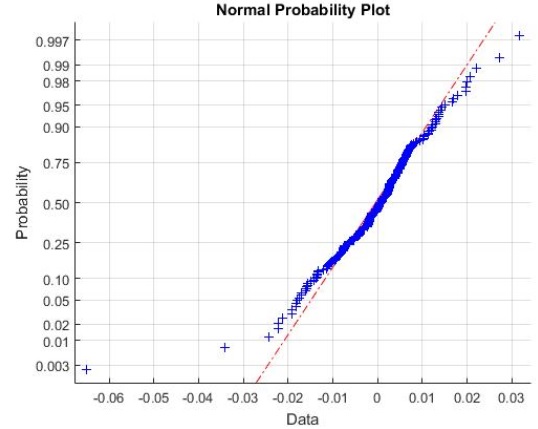}
		\caption{Daily Returns of BSE 200 for period 4 (06/15 - 04/16)}
		\label{fig:combined}
	\end{minipage}
\end{figure}
\FloatBarrier

\FloatBarrier
\begin{figure} [H]   
	\begin{minipage}[t]{0.45\textwidth}
		\includegraphics[width=\linewidth]{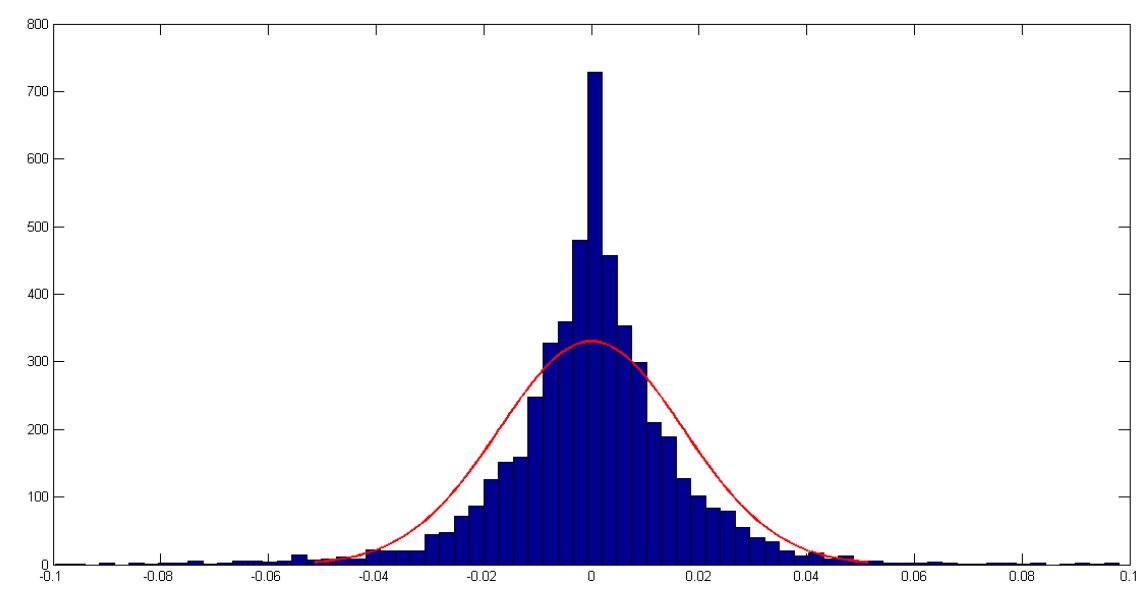}
		\caption{Histogram of Daily Returns for SSE Composite}
		\label{fig:immediate}
	\end{minipage}
	\hspace{\fill}
	\begin{minipage}[t]{0.45\textwidth}
		\includegraphics[width=\linewidth]{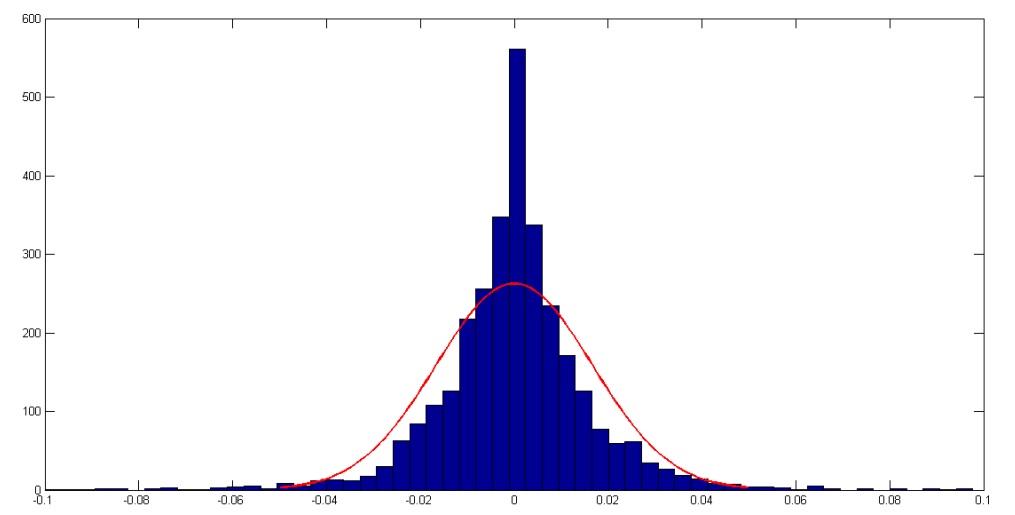}
		\caption{Histogram of Daily Returns of SSE Composite for period 1 (01/96 - 11/07) }
		\label{fig:proximal}
	\end{minipage}
	
	\vspace*{0.5cm} 
	\begin{minipage}[t]{0.45\textwidth}
		\includegraphics[width=\linewidth]{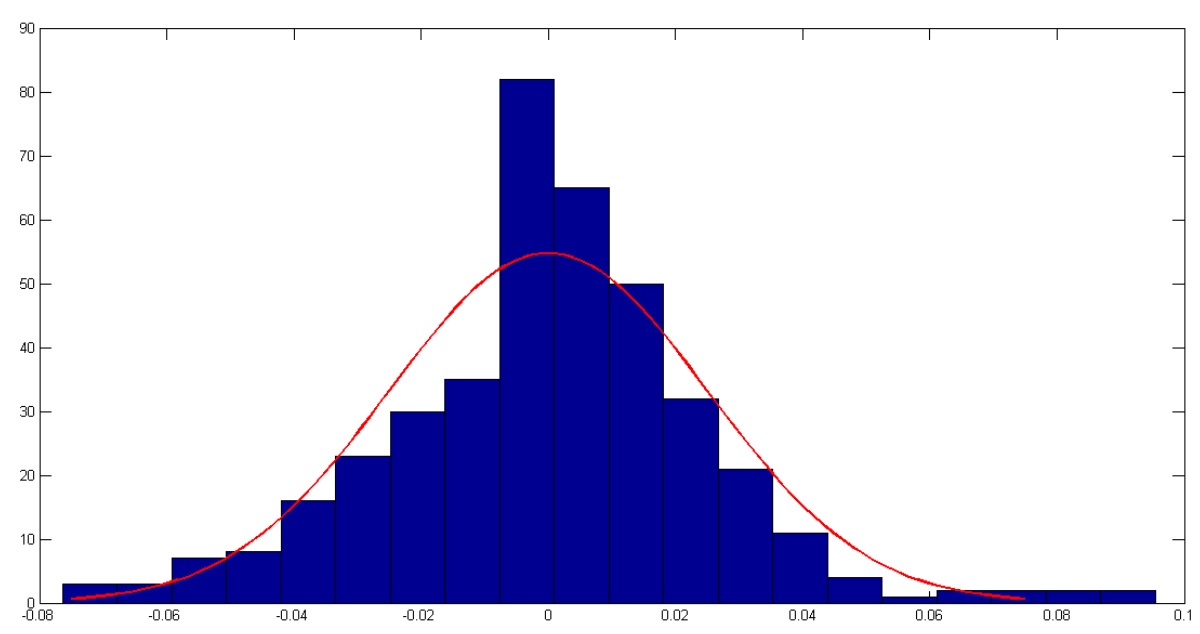}
		\caption{Histogram of Daily Returns of SSE Composite for period 2 (12/07 - 06/09) }
		\label{fig:distal}
	\end{minipage}
	\hspace{\fill}
	\begin{minipage}[t]{0.45\textwidth}
		\includegraphics[width=\linewidth]{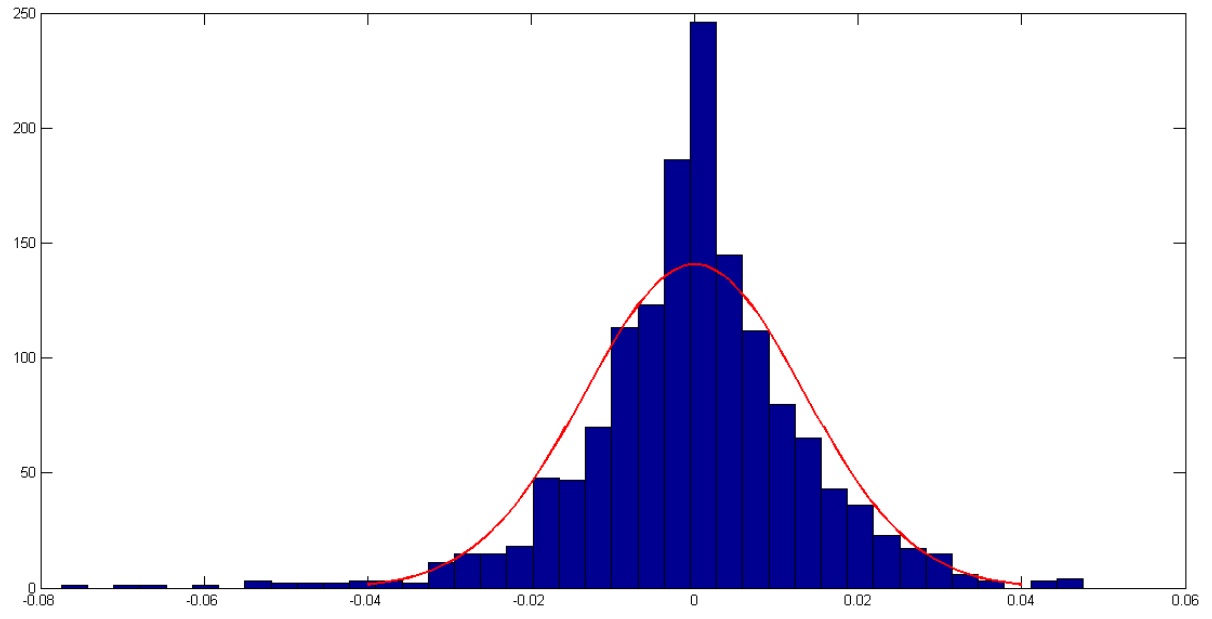}
		\caption{Histogram of Daily Returns of SSE Composite for period 3 (07/09 - 05/15) }
		\label{fig:combined}
	\end{minipage}
	\hspace{\fill}
	\begin{minipage}[t]{0.45\textwidth}
		\includegraphics[width=\linewidth]{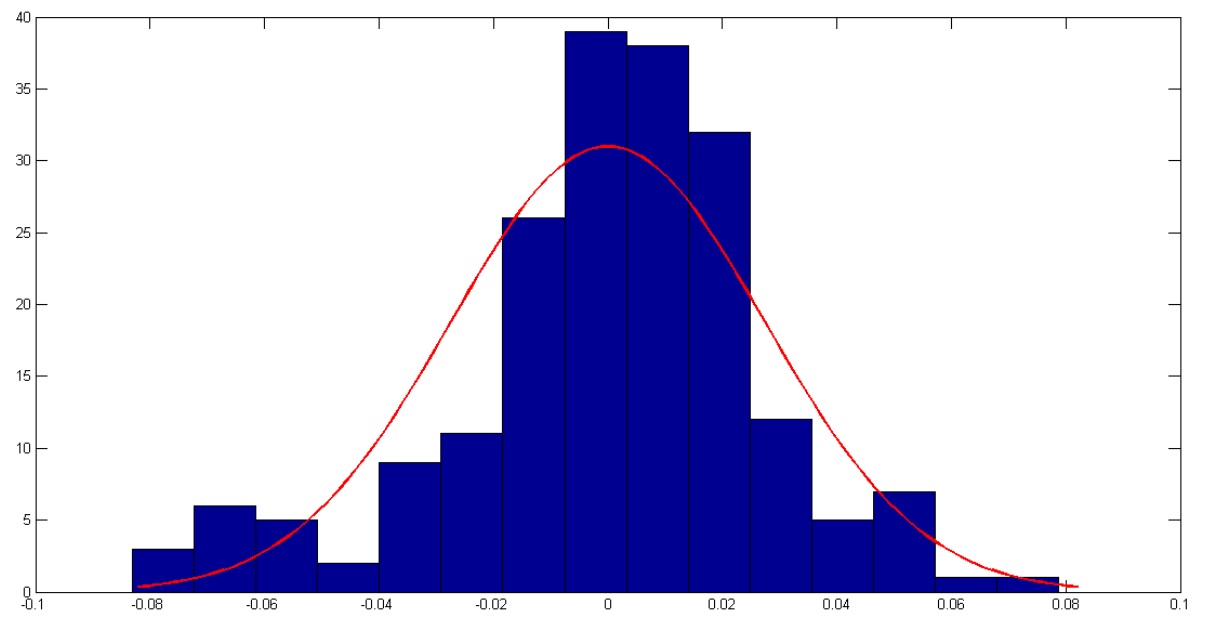}
		\caption{Histogram of Daily Returns of SSE Composite for period 4 (06/15 - 04/16)}
		\label{fig:combined}
	\end{minipage}
\end{figure}
\FloatBarrier

\FloatBarrier
\begin{figure} [H]   
	\begin{minipage}[t]{0.45\textwidth}
		\includegraphics[width=\linewidth]{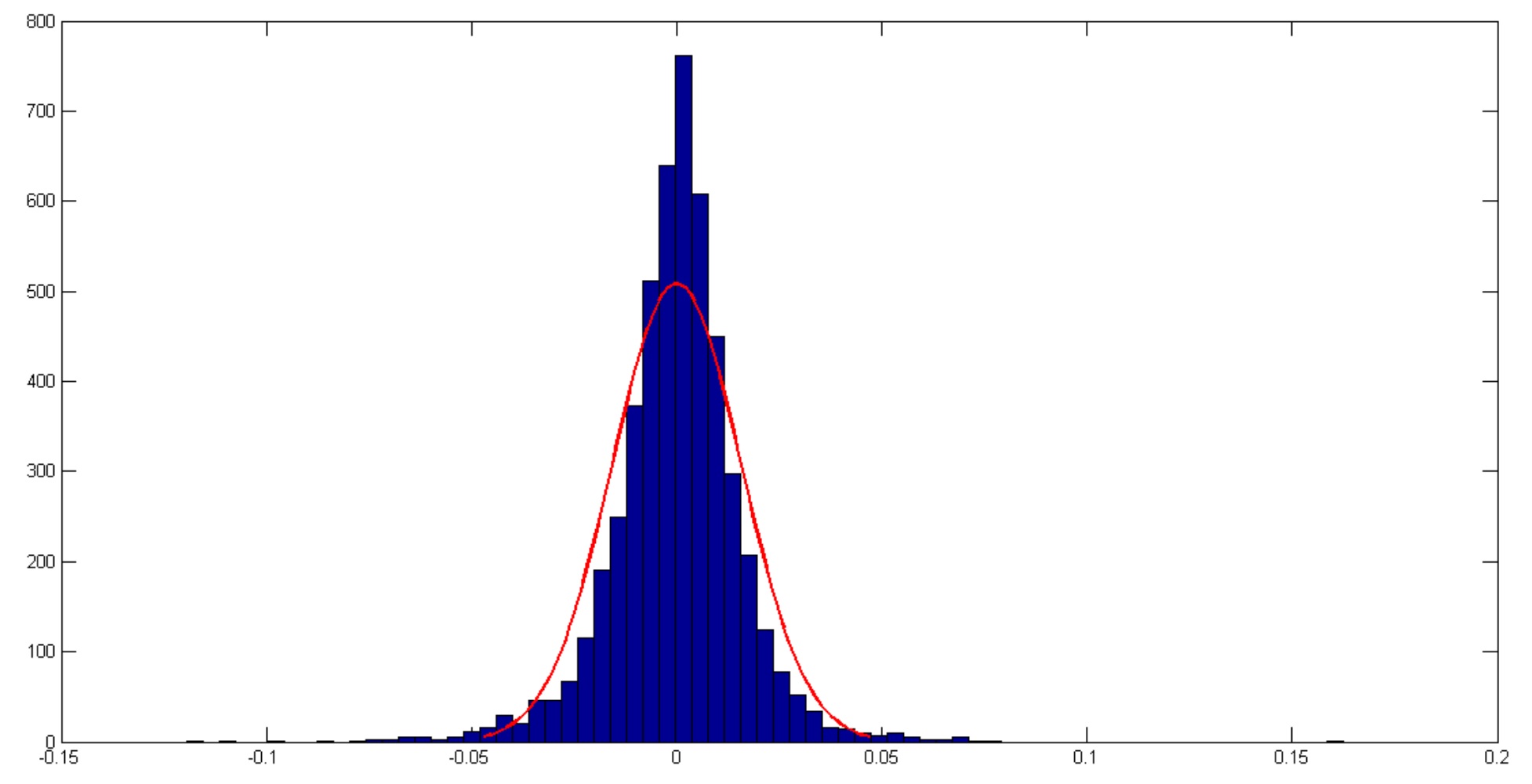}
		\caption{Histogram of Daily Returns for BSE 200}
		\label{fig:immediate}
	\end{minipage}
	\hspace{\fill}
	\begin{minipage}[t]{0.45\textwidth}
		\includegraphics[width=\linewidth]{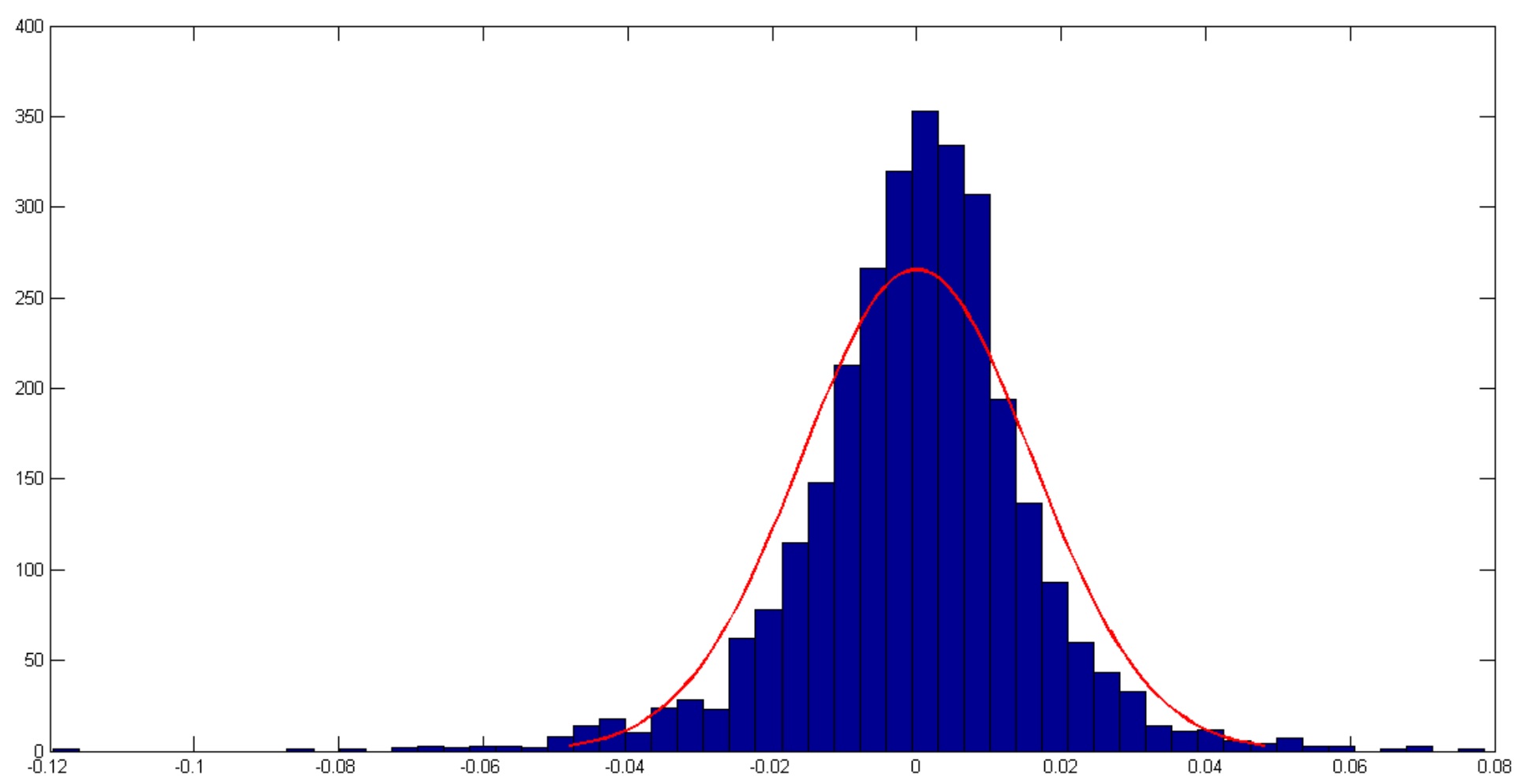}
		\caption{Histogram of Daily Returns of BSE 200 for period 1 (01/96 - 11/07) }
		\label{fig:proximal}
	\end{minipage}
	
	\vspace*{0.5cm} 
	\begin{minipage}[t]{0.45\textwidth}
		\includegraphics[width=\linewidth]{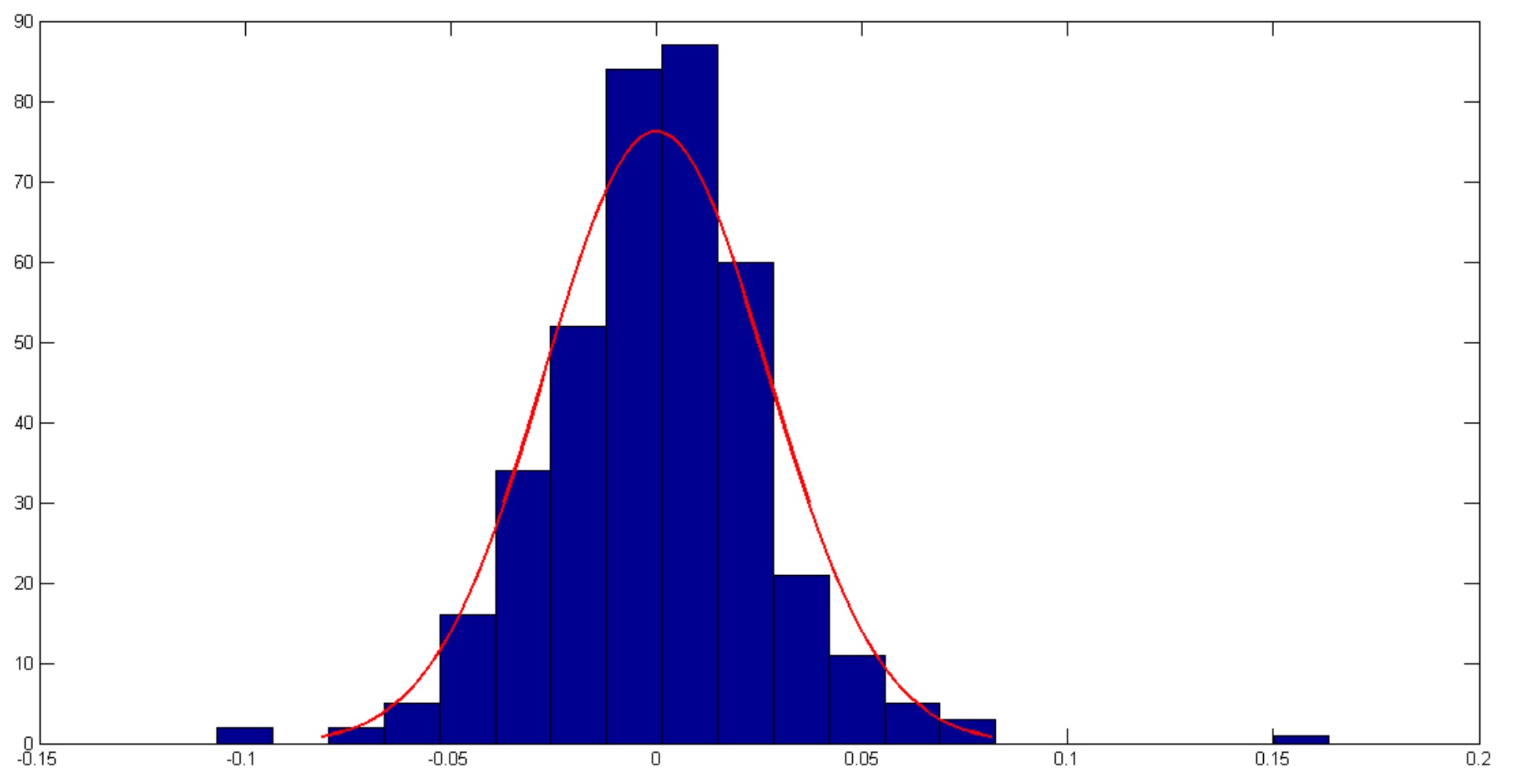}
		\caption{Histogram of Daily Returns of BSE 200 for period 2 (12/07 - 06/09) }
		\label{fig:distal}
	\end{minipage}
	\hspace{\fill}
	\begin{minipage}[t]{0.45\textwidth}
		\includegraphics[width=\linewidth]{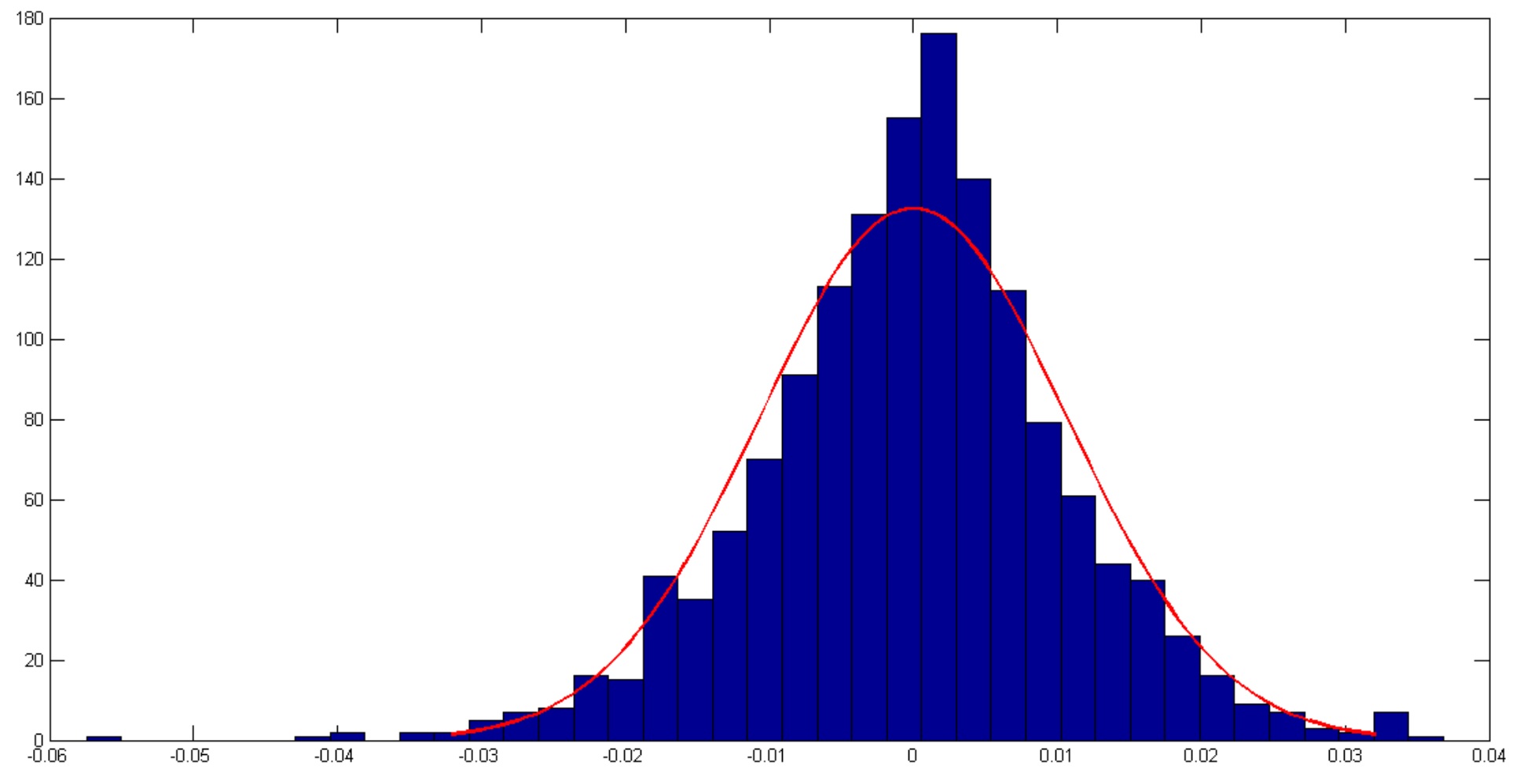}
		\caption{Histogram of Daily Returns of BSE 200 for period 3 (07/09 - 05/15) }
		\label{fig:combined}
	\end{minipage}
	\hspace{\fill}
	\begin{minipage}[t]{0.45\textwidth}
		\includegraphics[width=\linewidth]{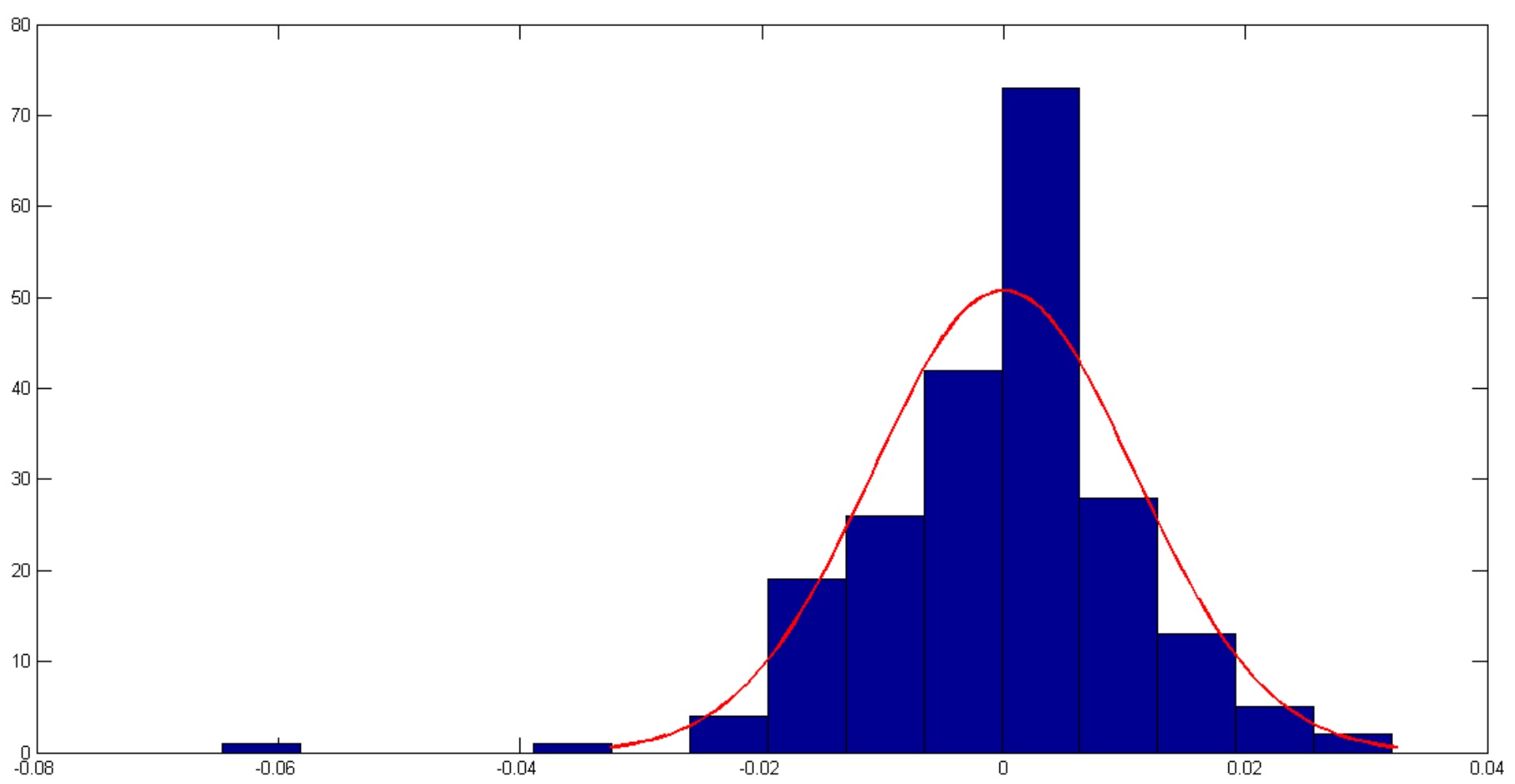}
		\caption{Histogram of Daily Returns of BSE 200 for period 4 (06/15 - 04/16)}
		\label{fig:combined}
	\end{minipage}
\end{figure}
\FloatBarrier
\noindent
Figure 3 - Figure 7 show the normal probability plots of the daily returns of SSE composite stock index. The normal probability plot plots the empirical cumulative distribution of the sample data versus the theoretical cumulative distribution function of a normal distribution. The horizontal axis plots the sorted sample data. The vertical axis plots the normal order statistic medians, calculated using the uniform order statistic medians and the inverse cumulative distribution function (icdf) of the normal distribution. If the sample data has a normal distribution, then the plot will appear linear. Distributions other than normal introduce curvature in the plot. 
\\
Many models in finance like the Black-Scholes model assume that the returns on indices or stock prices follow a log-normal distribution. This is a very crude assumption as we know that these returns follow fat tail distributions. It can be inferred that the daily returns of SSE Composite index for period 2 and period 4 to a very good approximation follow log-normal distribution. Similarly, for BSE 200 index, figure 8 - figure 12 show the normal probability plots of the daily returns. The daily returns of BSE 200 index for period 2 and period 4 to a very good approximation follow log-normal distribution. These were the periods of global financial distress. 
\\
\\
Figure 13 - Figure 22 show the histograms of the daily returns of both SSE composite index and BSE 200 index. These histograms have also been superimposed with a normal distribution curve to check whether the curve fits these histograms or not. It can be seen that the normal distribution curve does not fit the histogram for either of the two stock index and for any of the four periods.

\subsection{Runs Test}
The results of run test are given in the following tables; the first two tables give the values of run test for the entire 20 yr period of consideration and the four sub-periods. For SSE, it can be seen that Z values are more than + 1.96 overall and in period I, therefore, null hypothesis for the run test is rejected at 5 percent level of significance for the period before crisis. The run test for period-II, III, IV shows the sign of efficiency because the value of Z statistic is less than + 1.96 at 5 percent level of significance. Thus, the markets are expected to follow random walk. The state of efficiency has changed as the impact of recent financial crisis in the Chinese stock market. Hence, it can be concluded that it does not follow random walk over the time of study and for this reason SSE is considered to be weak form inefficient but the market has become efficient during and after the time period of the global financial crisis.
\\
\\
For the Indian Stock Market, i.e. the BSE it can be seen that Z values are more than + 1.96 overall and in periods before period IV which is the Chinese crisis, therefore the null hypothesis for the runs test is rejected at 5 percent level of significance for the period before crisis. The runs test for period IV however shows the sign of efficiency at 5\% significance level. Thus, the markets are expected to follow random walk. So, we can conclude from the results that the state of efficiency has changed as a result of the impact of recent Chinese crisis on the Indian stock market. Hence, it does not follow random walk over the time of study and for this reason BSE is considered to be weak form inefficient but the market is starting to become efficient with the onset of the Chinese crisis. 
\\
\\
It is interesting to note that throughout and in period-I, both the markets are inefficient but the global crisis makes the Chinese market efficient and Chinese crisis plays a role in making the Indian market efficient.

\FloatBarrier
\begin{table}[!htbp]
	\begin{adjustwidth}{0.0cm}{}
		\resizebox{\columnwidth}{!}{%
			\begin{tabular}{c|c|c|c|c|c} \hline
				& Daily & Period 1 (01/96-11/07) & Period 2 (12/07-06/09) & Period 3 (07/09-05/15) & Period 4 (06/15-04/16)\\ \hline
				N	&5153 	&3092	&399	&1465	&197\\ \hline
				Nruns & 2477 (2437) & 1459 (1431) & 192 (189) &720 (712) & 106 (106)\\ \hline
				$n_1$ & 2488 (2591) & 1432 (1537) & 211 (192) & 732 (760) & 84 (102)\\ \hline
				$n_0$ & 2665 (2320) & 1660 (1356) & 188 (192) & 733 (677) & 113 (95)\\ \hline
				$n_2$ & 0 (242) & 0 (199) & 0 (15) & 0 (28) & 0 (0)\\ \hline
				Z & 2.7049 (0.3299*) & 2.8608 (0.3860*) & 0.7380* (0.3577*) & 0.6795* (0.2438*) & 1.1880* (0.8760*)\\ \hline
				p-value & 0.0068 (0.7415) & 0.0042 (0.6995) & 0.4605 (0.7205) & 0.4968 (0.8703) & 0.234 (0.3812)\\ \hline
			\end{tabular}%
		}
		\caption{Runs Tests for the returns of the SSE composite stock index relative to mean (zero)}\label{tbl:names}
	\end{adjustwidth}
\end{table}
\FloatBarrier

Test data, returned as a structure with the following fields.
\begin{itemize}
	\item nruns — The number of runs
	\item n1 — The number of values above v
	\item n0 — The number of values below v
	\item n2 — the number of values equal to v
	\item z — The test statistic
\end{itemize}
where, v is either the mean value or any arbitrary number which is 0 in our case. 
\\
The test is based on the number of runs of consecutive values above or below the mean of x (zero). The result h is 1 if the test rejects the null hypothesis that the data is in random order at the 5\% significance level, or 0 otherwise.

\FloatBarrier
\begin{table}[!htbp]
	\begin{adjustwidth}{0.0cm}{}
		\resizebox{\columnwidth}{!}{%
			\begin{tabular}{c|c|c|c|c|c} \hline
				& Daily & Period 1 (01/96-11/07) & Period 2 (12/07-06/09) & Period 3 (07/09-05/15) & Period 4 (06/15-04/16)\\ \hline
				N&5032 &2965&	383	&1471&	215\\ \hline
				Nruns	&2214 (2210)	&1294 (1292)	&172 (170)	&651 (653)	&98 (98)\\ \hline
				$n_1$	&2742 (2638)	&1636 (1550)&	193	(198)& 796 (772) &	117 (120)\\ \hline
				$n_0$ &	2287 (2393)&	1326 (1414)&190 (185)&	674 (698)&97 (94)
				\\ \hline
				$n_2$	&3 (1)&	3 (1)&	0 (0)&	1 (1)&	1 (1)\\ \hline
				Z	&7.9746 (8.481)	&6.3650 (6.8993)	&2.0455 (2.2312)&	4.1740 (4.2185)	&1.1842* (1.1018*)\\ \hline
				p-value	&1.4802e-15 (2.072e-17)&	1.9222e-10 (4.93e-12)&	0.0407 (0.0256)&	2.9677e-05 (2.4233e-05)&	0.2363 (0.2704)\\ \hline
			\end{tabular}%
		}
		\caption{Runs Tests for the returns of the BSE 200 stock index relative to mean (zero)}\label{tbl:names}
	\end{adjustwidth}
\end{table}
\FloatBarrier

The test is based on the number of runs of consecutive values above or below the mean of x (zero). The result h is 1 if the test rejects the null hypothesis that the data is in random order at the 5\% significance level, or 0 otherwise.

\subsection{Augmented Dickey-Fuller (ADF) Test}

The results of ADF Augmented Dickey-Fuller Test of unit root for all the periods are given in the table below for both the markets. The more negative the DF statistic value is and lesser is the p-value, more chance are that the series is stationary. So, we expect the value of the DF statistic to be close to zero and the p value to be farther away from zero. A p-value of 0.01 here means that there is $1\%$ chance that stationarity of a process cannot be rejected which implies $1\%$ chance that non-stationarity is possible. So, there is a $99\%$ chance that the process is stationary. The number of lags is basically the default lag order of an ARMA process. 
\\
\\
The values in the table have high negative DF statistic values and p-value equal to 0.01 for all the periods for both the BSE and SSE composite indexes indicating that they are stationary series. This indicates that both the markets don’t follow random walks and are weak form inefficient.

\FloatBarrier
\begin{table}[!htbp]
	\begin{adjustwidth}{0.0cm}{}
		\resizebox{\columnwidth}{!}{%
			\begin{tabular}{c|c|c|c|c|c} \hline
				& Daily & Period 1 (01/96-11/07) & Period 2 (12/07-06/09) & Period 3 (07/09-05/15) & Period 4 (06/15-04/16)\\ \hline
				ADF Test Statistic&-15.22 &-12.66&	-7.40	& -10.57 &	-6.09\\ \hline
				p-value & 0.01 	& 0.01	& 0.01	& 0.01	& 0.01\\ \hline
				Number of Lags& 17	& 14 & 7 & 11 & 5\\ \hline
				number of Observations & 5153 & 3092 & 399 & 1465 & 197
				\\ \hline
			\end{tabular}%
		}
		\caption{ADF Test for the returns of the SSE composite stock index}\label{tbl:names}
	\end{adjustwidth}
\end{table}
\FloatBarrier

\FloatBarrier
\begin{table}[!htbp]
	\begin{adjustwidth}{0.0cm}{}
		\resizebox{\columnwidth}{!}{%
			\begin{tabular}{c|c|c|c|c|c} \hline
				& Daily & Period 1 (01/96-11/07) & Period 2 (12/07-06/09) & Period 3 (07/09-05/15) & Period 4 (06/15-04/16)\\ \hline
				ADF Test Statistic&-15.42 &-12.86&	-6.35	& -11.35 &	-6.94\\ \hline
				p-value & 0.01 	& 0.01	& 0.01	& 0.01	& 0.01\\ \hline
				Number of Lags& 17	& 14 & 7 & 11 & 5\\ \hline
				number of Observations & 5032 & 2965 & 383 & 1471 & 215
				\\ \hline
			\end{tabular}%
		}
		\caption{ADF Test for the returns of the SSE 200 stock index}\label{tbl:names}
	\end{adjustwidth}
\end{table}
\FloatBarrier

\subsection{Auto Correlation Function (ACF)}
The ACF is used to determine the independence of stock price changes. It measures the amount of linear dependence between observations in time series that are separated by lag. The ACF and t-values of BSE and SSE for all the periods are presented in the following tables.
\\
\\
If t $<$ -1.96 or t $>$ 1.96, we reject the null hypothesis at 95$\%$ confidence interval that the series is weak form efficient. According to the t-values of the coefficients, the values for SSE are found to be significant at three lags overall and at six, five, four and six lags respectively in the four periods respectively. The more insignificant the values, the more strongly we can reject the null that the market follows a random walk. Thus, we reject the null hypothesis that the market is weakly efficient. But we could also point out the fact that the market efficiency seems to be increasing for the last period after the fall in inefficiency during the recession period.  
\\
\\
The behaviour of BSE in this period is found to be way better overall as compared to SSE because of significance at nine lags overall. The t-values also rejects the null hypothesis of market efficiency at seven, three, nine and four lags respectively. The market efficiency turned out to be better in the post-recession period, even more than the pre-crisis period but fall after the start of the Chinese crisis. Therefore, the ACFs for BSE and SSE are highly autocorrelated and it can be concluded that both BSE and SSE, both are considered to especially be inefficient markets during the crisis periods.
\\
\\
Therefore, as the markets are less inefficient there is possibility of earning extra income on the account of market inefficiency.

(Here, h = 1 indicates the rejection of the null hypothesis that the residuals are not autocorrelated. pValue indicates the strength at which the test rejects the null hypothesis. Whenever the p-value is less than 0.01, there is strong evidence to reject the null hypothesis that the residuals are not autocorrelated meaning that market is weak form efficient.)
\\
\\
From the Ljung box Q-statistic test, we find that the Indian Stock market is starting to follow random walk after the recession period and same is the case for the Chinese stock market, with an exception for the last period of Chinese crisis where it moves away from efficiency.

\FloatBarrier
\begin{table}[!htbp]
	\begin{adjustwidth}{0.0cm}{}
		\resizebox{\columnwidth}{!}{%
			\begin{tabular}{c|c|c|c|c|c} \hline
				Lag & Daily & Period 1 (01/96-11/07) & Period 2 (12/07-06/09) & Period 3 (07/09-05/15) & Period 4 (06/15-04/16)\\ \hline
				1 &	0.0319 (3.57) &	0.0039 (0.34)	&	0.0063 (0.59)& 0.0375 (5.47)	 &	0.0893 (3.41)\\ \hline
				2 &	-0.0505 (-5.65) &	-0.0399 (-3.47) &	-0.0421 (-3.97) &	0.0035 (0.51) &	-0.1169 (-4.46)\\ \hline
				3 &	0.0576 (6.45) &	0.074 (6.44)	&0.0706 (6.66) &	0.0146 (2.13) &	0.0587 (2.24)\\ \hline
				4 &	0.083 (9.3)	& 0.0545 (4.74)	& 0.0872 (8.23) &	0.016 (2.33)	& 0.1604 (6.12)\\ \hline
				5 &	0.0057 (0.64)	&0.0011 (0.096) &	-0.0199 (-1.88) &	-0.0401 (-5.84) &	0.0706 (2.67)\\ \hline
				6 &	-0.0555 (-6.21) &	-0.0546 (-4.75) &	-0.0209 (-1.97) &	-0.005 (-0.73) &	-0.1753 (-6.69)\\ \hline
				7 &	0.0184 (2.06) &	0.0048 (0.42) &	0.0226 (2.13) &	0.0383 (5.58) &	-0.0059 (-0.23)\\ \hline
				8 &	0.0256 (2.87) &	-0.0561 (-4.88) &	-0.0265 (-2.50) &	0.0242 (3.53) &	0.1894 (7.23)\\ \hline
				9 &	-0.0224 (-2.50) &	-0.0294 (-2.56) &	-0.0531 (-5.01) &	0.016 (2.33) &	-0.046 (-1.75)\\ \hline
				10 &	-0.0206 (-2.31) &	0.0491 (4.27) &	-0.0389 (-3.67) &	0.0887 (12.94) &	-0.2532 (-9.66)\\ \hline
				11 &	0.0203 (2.27) &	0.0992 (8.64) &	0.0391 (3.69) &	-0.0388 (-5.66) &	-0.1026 (-3.92)\\ \hline
				12 &	0.0405 (4.53) &	0.0762 (6.63) &	0.0228 (2.15) &	-0.0117 (-1.70) &	0.0333 (1.27)\\ \hline
				13 &	0.0637 (7.13) &	0.0297 (2.59) &	0.0839 (7.92) &	0.0199 (2.90) &	0.1125 (4.30)\\ \hline
				14 &	-0.0263 (-2.94) &	0.0137 (1.19) &	0.0093 (0.88) &	-0.0376 (-5.48) &	-0.1546 (-5.90)\\ \hline
				15 &	0.0635 (7.10) &	0.1124 (9.78)&	0.0173 (1.63) &	0.0268 (3.91) &	-0.0192 (-0.73)\\ \hline
				16 &	0.0225 (2.51) &	0.0355 (3.09) &	-0.0643 (-6.06)&	0.0322 (4.69)&	0.0579 (2.21)\\ \hline
				17 &	-0.0055 (-0.61)&	-0.0137 (-1.19) &	-0.0318 (-3.00)&	0.0338 (4.93) &	0.006 (0.23)\\ \hline
				18 &	0.0139 (1.56) &	0.0304 (2.65)&	0.0515 (4.86) &	0.0213 (3.10) &	-0.0372 (-1.42)\\ \hline
				19 &	-0.0484 (-5.41) &	-0.0645 (-5.62) &	-0.0185 (-1.75)&	-0.003 (-0.44)&	-0.0688 (-2.63)\\ \hline
				20 &	0.0215 (2.40)	& -0.0026 (-0.23)&	-0.0656 (-6.19)&	0.0303 (4.42)&	0.1203 (4.59)\\ \hline
				Standard
				Deviation &	0.039949 &	0.051367 &	0.047381 &	0.030664 &	0.117175\\ \hline
				Standard
				Error &	0.008933 &	0.011486 &	0.010595 &	0.006857 &	0.026201\\ \hline
				Ljung Box Q-Stat &	89.7187* &	82.9163*& 10.5024 & 24.2587 & 37.551*\\ \hline
				p-value &	8.30E-11 &	1.25E-09 &	0.9581 &	0.2313 &	0.01\\ \hline
			\end{tabular}%
		}
		\caption{Serial Correlation Coefficients for Returns of the SSE composite stock index (corresponding t-values)}\label{tbl:names}
	\end{adjustwidth}
\end{table}
\FloatBarrier

\FloatBarrier
\begin{table}[!htbp]
	\begin{adjustwidth}{0.0cm}{}
		\resizebox{\columnwidth}{!}{%
			\begin{tabular}{c|c|c|c|c|c} \hline
				Lag & Daily & Period 1 (01/96-11/07) & Period 2 (12/07-06/09) & Period 3 (07/09-05/15) & Period 4 (06/15-04/16)\\ \hline
				1	& 0.1059 (14.19) & 	0.1326 (10.35) & 	0.1066 (7.54) & 	0.1017 (15.11) & 	0.0529 (3.48)\\ \hline
				2 & 	0.0044 (0.59) & 	-0.0553 (-4.32) & 	0.0407 (2.81) & 	-0.0001 (-0.01) & 	0.0231 (1.52) \\ \hline
				3 & 	-0.0005 (-0.07)& 	-0.0011 (-0.08) & 	0.0001 (0.006) & 	-0.0106 (-1.58) & 	0.0229 (1.51)\\ \hline
				4 & 	-0.0291 (-3.90) & 	0.0405 (3.16)& 	-0.0896 (-6.18)& 	-0.0194 (-2.88)& 	-0.0543 (-3.58)\\ \hline
				5 & 	-0.0351 (-4.70)& 	-0.0092 (-0.72)& 	-0.0752 (-5.19) & 	-0.0022 (-0.33)& 	-0.108 (-7.11)\\ \hline
				6 & 	-0.0319 (-4.28)& 	-0.0744 (-5.81)& 	-0.0309 (-2.13)& 	-0.0036 (-0.54)& 	-0.0716 (-4.72)\\ \hline
				7 & 	0.0329 (4.41)& 	-0.0397 (-3.09)& 	0.0944 (6.51)& 	-0.0028 (-0.42)& 	0.0854 (5.63)\\ \hline
				8 & 	0.0309 (4.14)& 	0.024 (1.87)& 	0.1192 (8.22) & 	-0.0169 (-2.51)& 	-0.0391 (-2.58)\\ \hline
				9 & 	0.0331 (4.44)& 	0.0686 (5.36)& 	0.0294 (2.03)& 	0.0062 (0.92)& 	0.0906 (5.97)\\ \hline
				10 & 	0.0117 (1.57)& 	0.0811 (6.33)& 	-0.0057 (-0.39)& 	-0.0373 (-5.54)& 	0.1132 (7.46)\\ \hline
				11 & 	-0.0333 (-4.46)& 	-0.0334 (-2.60) & 	-0.0294 (-2.0)& 	-0.0365 (-5.42)& 	0.0134 (0.88)\\ \hline
				12 & 	0.0073 (0.98)& 	0.0269 (2.10)& 	-0.0085 (-0.59)& 	0.0226 (3.36)& 	-0.0649 (-4.28)\\ \hline
				13 & 	-0.0003 (-0.04)& 	0.0882 (6.88)& 	-0.0309 (-2.13)& 	-0.0112 (-1.66)& 	-0.0877 (-5.78)\\ \hline
				14 & 	0.0453 (6.07)& 	0.0622 (4.85)& 	0.0966 (6.67)& 	0.0248 (3.69) & 	-0.0381 (-2.50)\\ \hline
				15 & 	0.0095 (1.27)& 	0.0138 (1.07)& 	0.0407 (2.80)& 	-0.0077 (-1.14)& 	-0.0816 (-5.38)\\ \hline
				16 & 	0.0022 (0.29)& 	0.0024 (0.19)& 	0.0623 (4.29)& 	-0.0254 (-3.78)& 	-0.0939 (-6.18)\\ \hline
				17 & 	0.0279 (3.74)& 	0.0125 (0.98)& 	0.071 (4.89)& 	0.023 (3.41)& 	-0.049 (-3.22)\\ \hline
				18 & 	-0.0031 (-0.42)& 	-0.0131 (-1.02)& 	-0.0553 (-3.81) & 	0.0201 (2.99)& 	0.0496 (3.27)\\ \hline
				19 & 	-0.0116 (-1.55) & 	-0.0549 (-4.29)& 	-0.0571 (-3.94)& 	0.0216 (3.21)& 	0.0087 (0.57)\\ \hline
				20 & 	-0.0196 (-2.63)& 	-0.0767 (-5.99)& 	-0.0514 (-3.55)& 	-0.0051 (-0.76)& 	0.0434 (2.85)\\ \hline
				Standard
				Deviation & 	0.033358 & 	0.057283 & 	0.064832 & 	0.030089 & 	0.067885\\ \hline
				Standard
				Error & 	0.007459 & 	0.012809 & 	0.014497 & 	0.006728 & 	0.01518\\ \hline
				Ljung Box Q-Stat &	128.45* &	107.51* & 24.16 & 24.62 & 19.97\\ \hline
				p-value &	0.00 &	5.56E-14 &	0.2355 &	0.2162 &	0.4598 \\ \hline
			\end{tabular}%
		}
		\caption{Serial Correlation Coefficients for Returns of the BSE 200 stock index (corresponding t-values)}\label{tbl:names}
	\end{adjustwidth}
\end{table}
\FloatBarrier
\noindent

\FloatBarrier
\begin{figure} [h]   
	\begin{minipage}[t]{0.45\textwidth}
		\includegraphics[width=\linewidth]{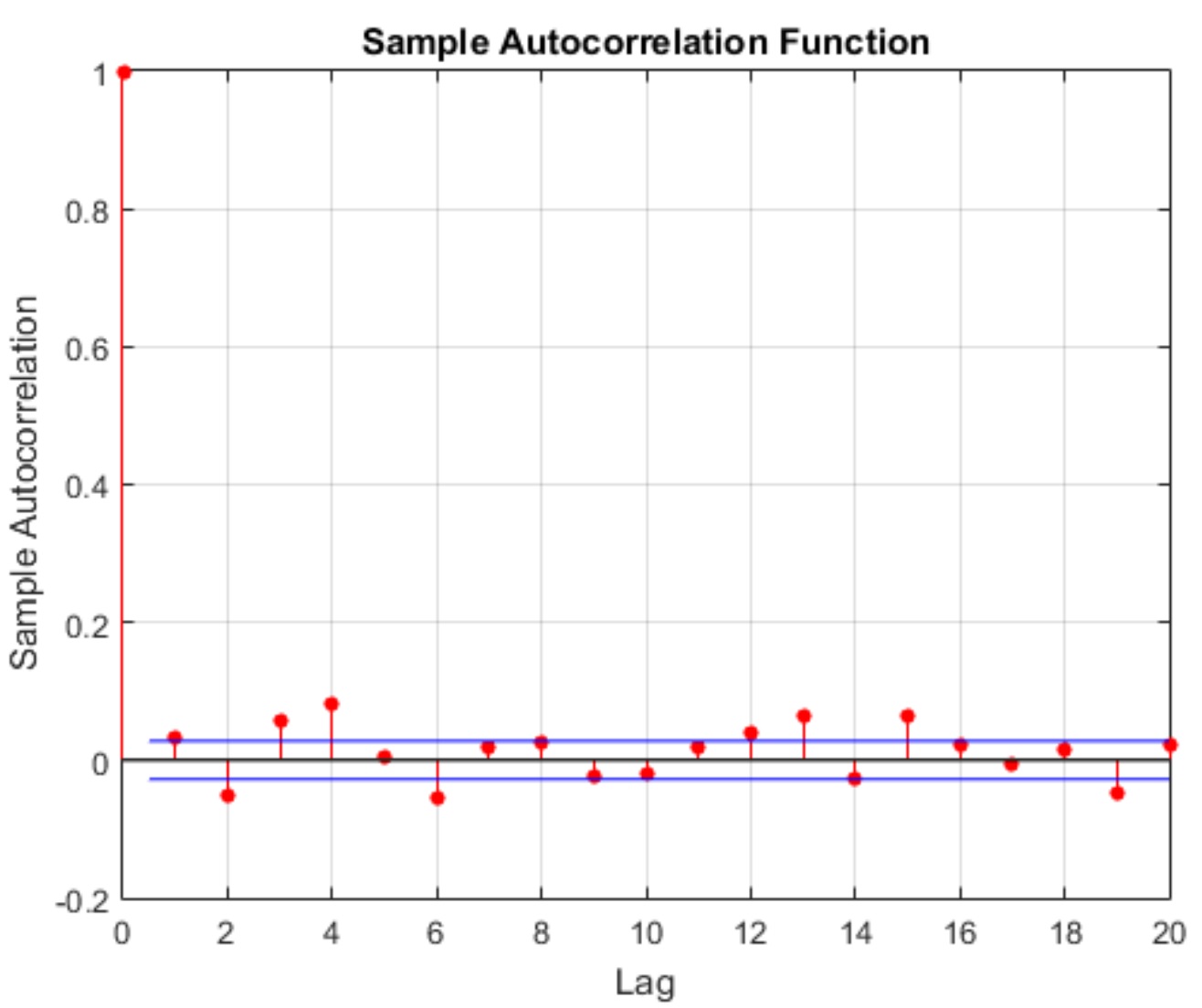}
		\caption{ACF coefficients for the period 01/96 - 04/16 for SSE}
		\label{fig:immediate}
	\end{minipage}
	\hspace{\fill}
	\begin{minipage}[t]{0.45\textwidth}
		\includegraphics[width=\linewidth]{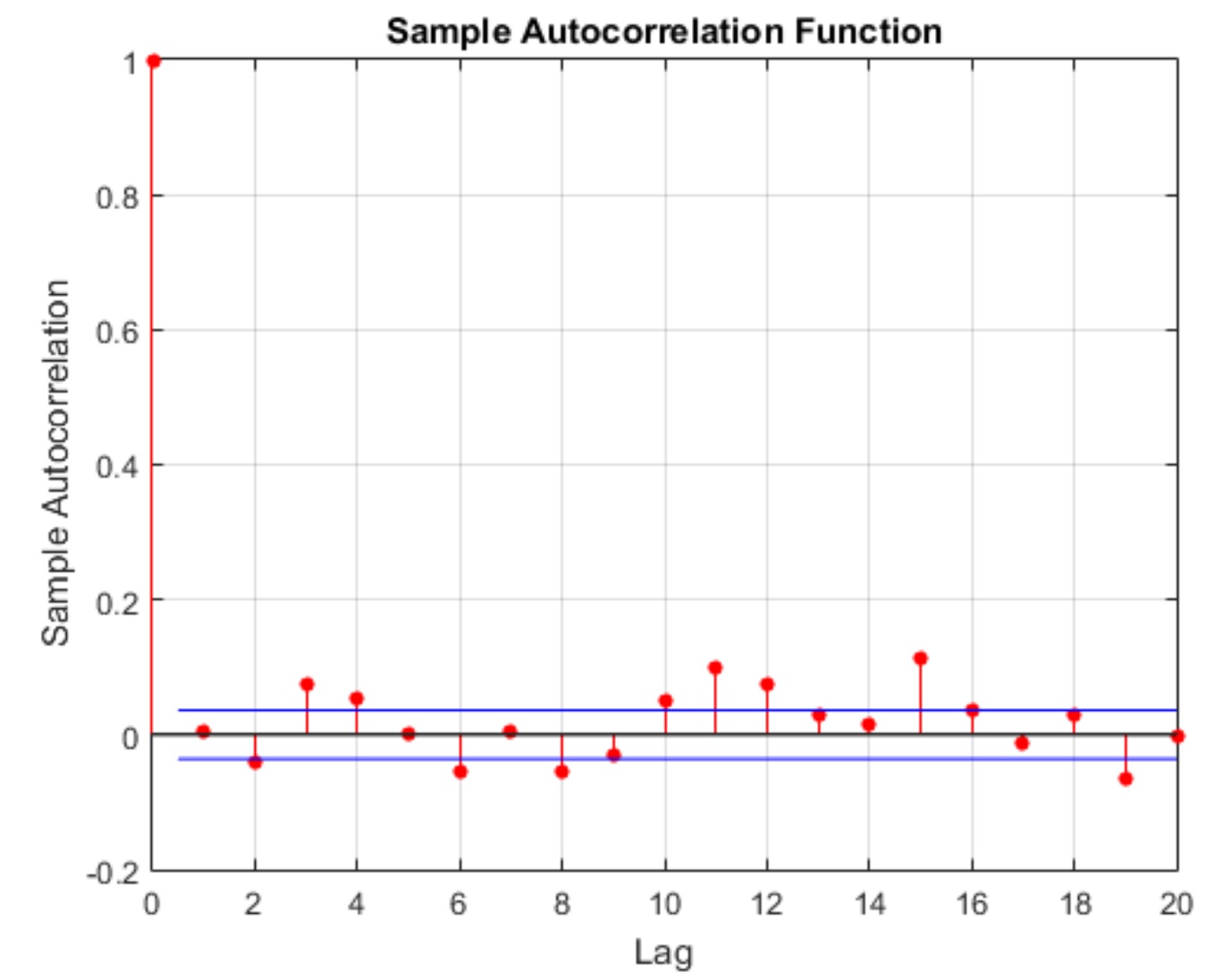}
		\caption{ACF coefficients for Period 1 (01/96 - 11/07) for SSE}
		\label{fig:proximal}
	\end{minipage}
	
	\vspace*{0.5cm} 
	\begin{minipage}[t]{0.45\textwidth}
		\includegraphics[width=\linewidth]{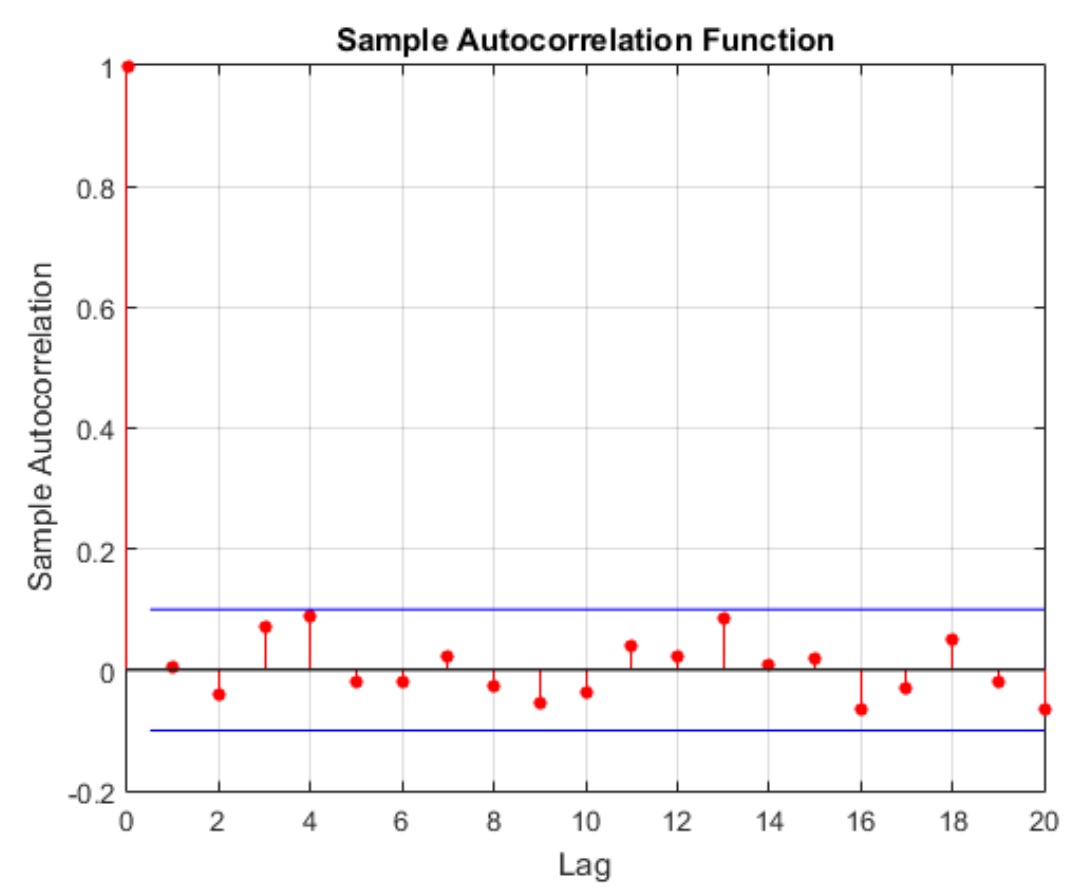}
		\caption{ACF coefficients for Period 2 (12/07 - 06/09) for SSE}
		\label{fig:distal}
	\end{minipage}
	\hspace{\fill}
	\begin{minipage}[t]{0.45\textwidth}
		\includegraphics[width=\linewidth]{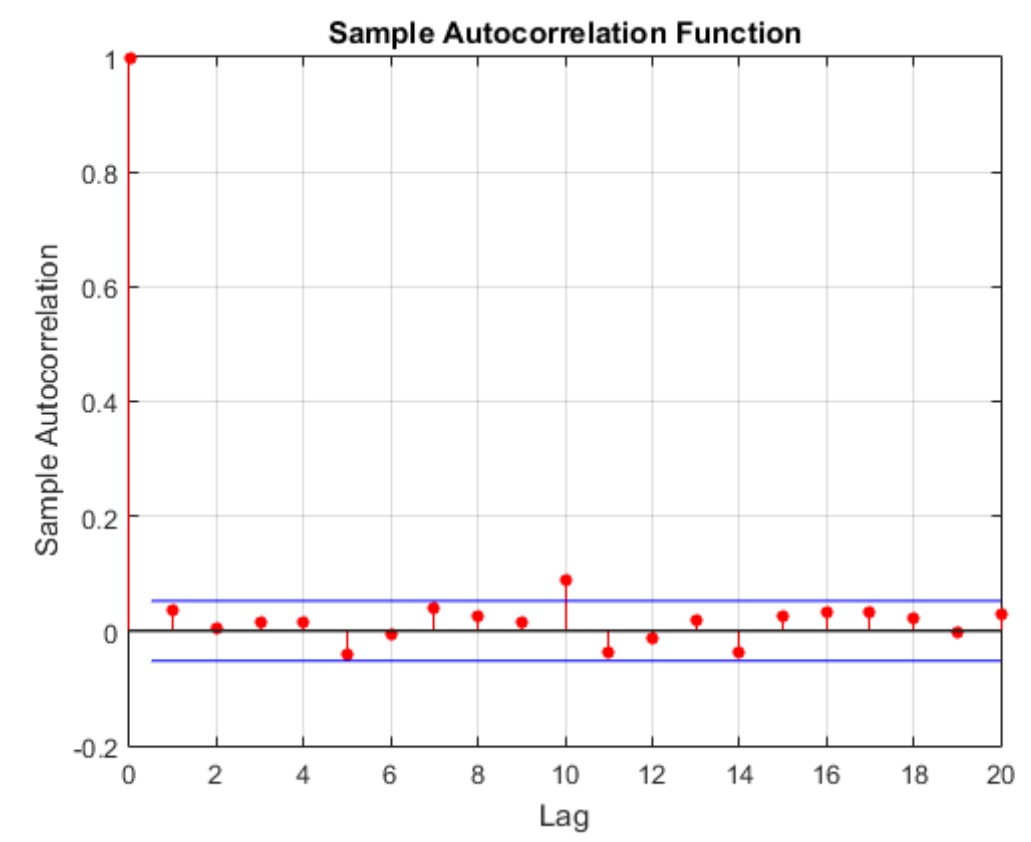}
		\caption{ACF coefficients for Period 3 (07/09 - 05/15) for SSE}
		\label{fig:combined}
	\end{minipage}
	\hspace{\fill}
	\begin{minipage}[t]{0.45\textwidth}
		\includegraphics[width=\linewidth]{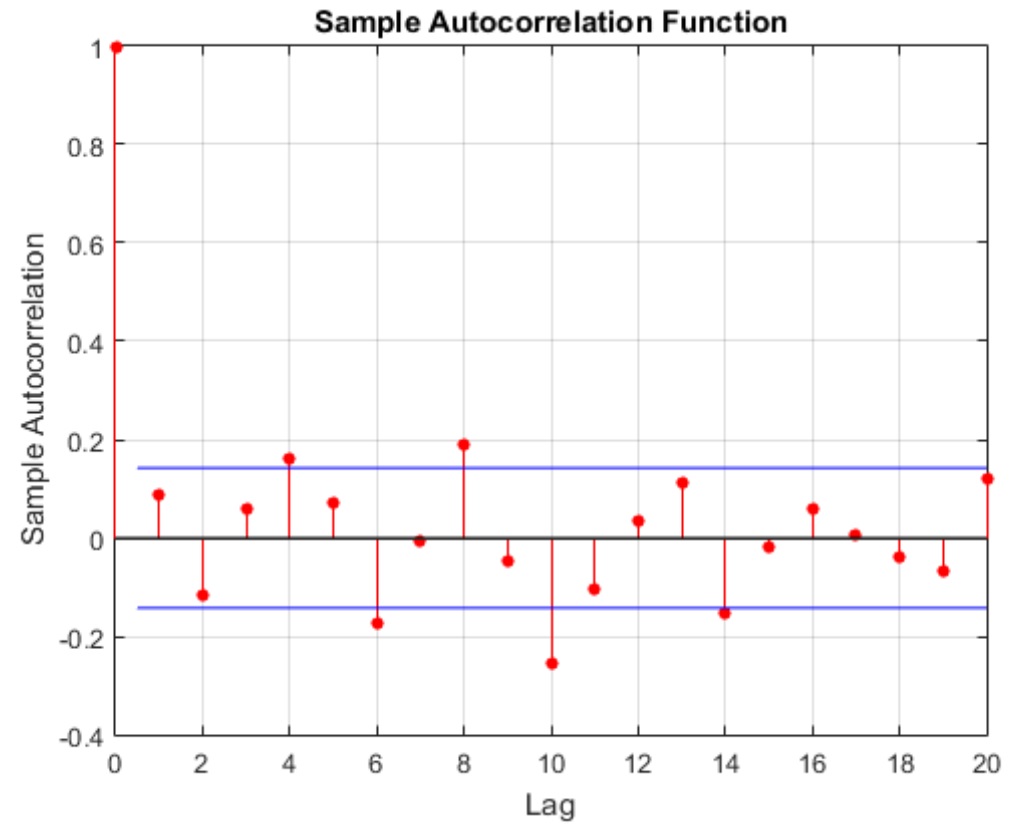}
		\caption{ACF coefficients for Period 4 (06/15 - 04/16) for SSE}
		\label{fig:combined}
	\end{minipage}
\end{figure}
\FloatBarrier

\FloatBarrier
\begin{figure} [h]   
	\begin{minipage}[t]{0.45\textwidth}
		\includegraphics[width=\linewidth]{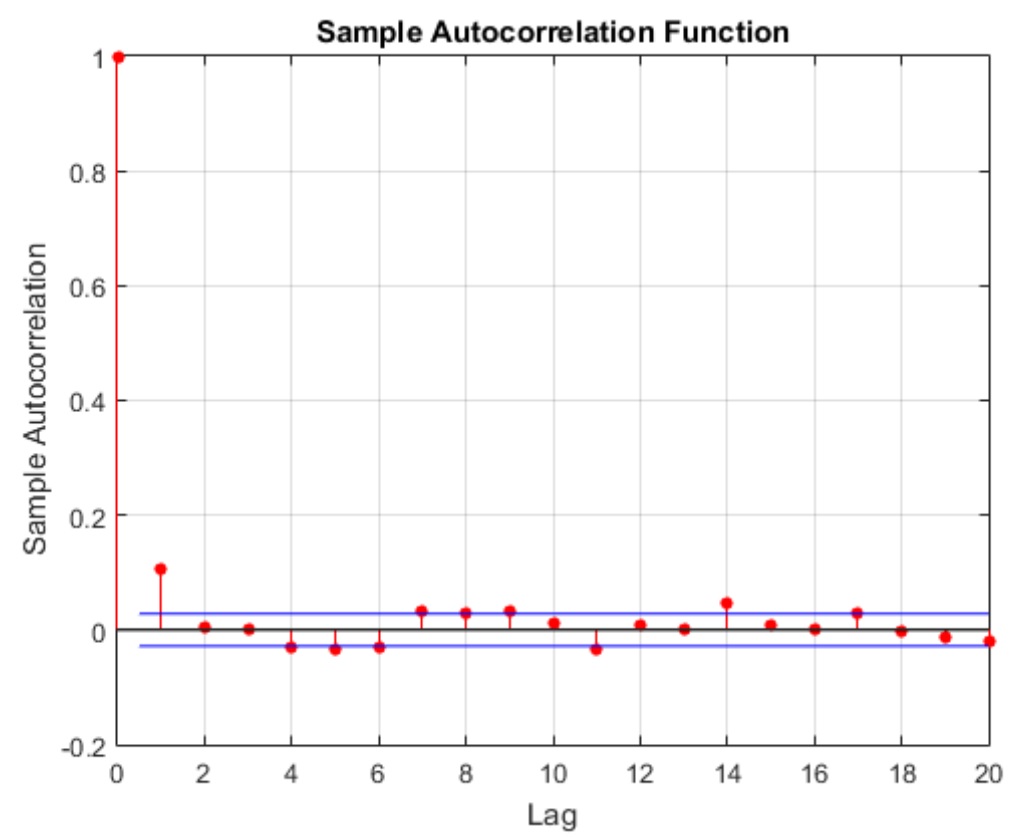}
		\caption{ACF coefficients for the period 01/96 - 04/16 for BSE}
		\label{fig:immediate}
	\end{minipage}
	\hspace{\fill}
	\begin{minipage}[t]{0.45\textwidth}
		\includegraphics[width=\linewidth]{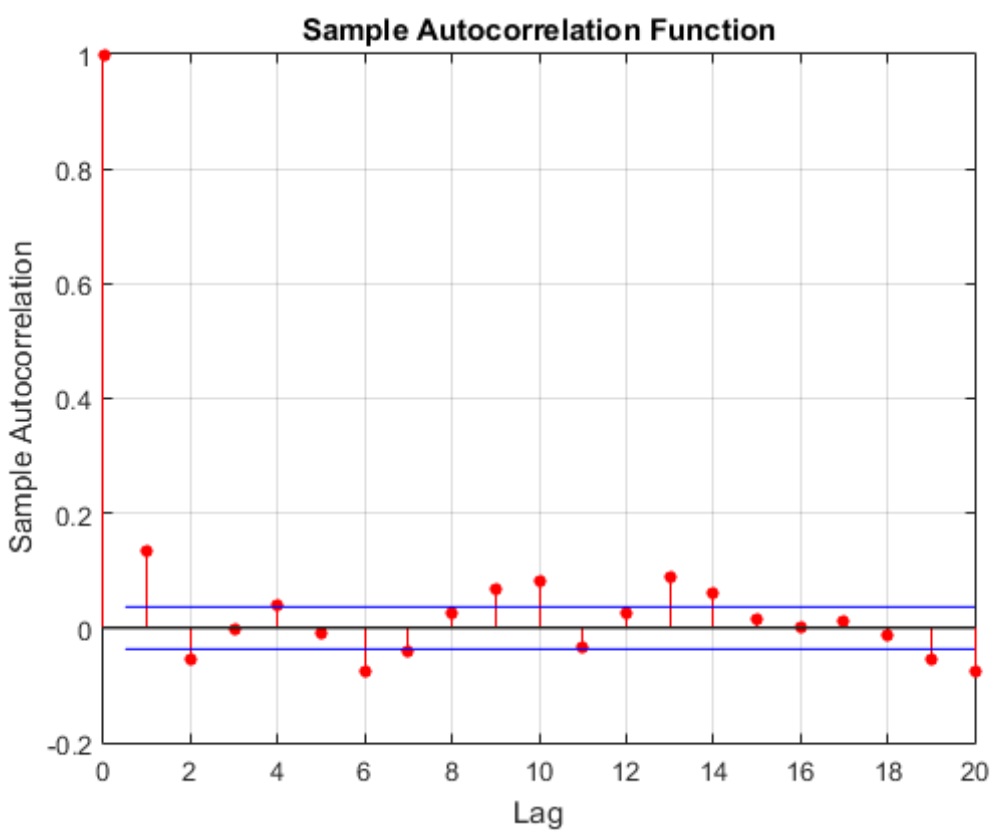}
		\caption{ACF coefficients for Period 1 (01/96 - 11/07) for BSE}
		\label{fig:proximal}
	\end{minipage}
	
	\vspace*{0.5cm} 
	\begin{minipage}[t]{0.45\textwidth}
		\includegraphics[width=\linewidth]{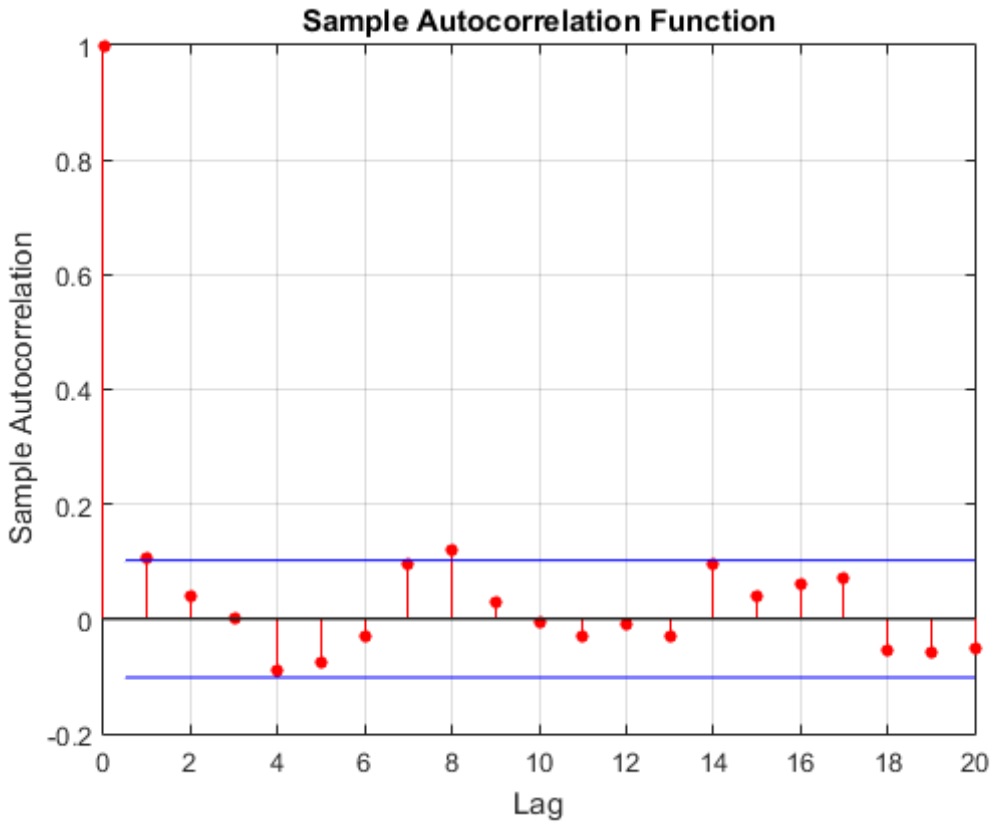}
		\caption{ACF coefficients for Period 2 (12/07 - 06/09) for BSE}
		\label{fig:distal}
	\end{minipage}
	\hspace{\fill}
	\begin{minipage}[t]{0.45\textwidth}
		\includegraphics[width=\linewidth]{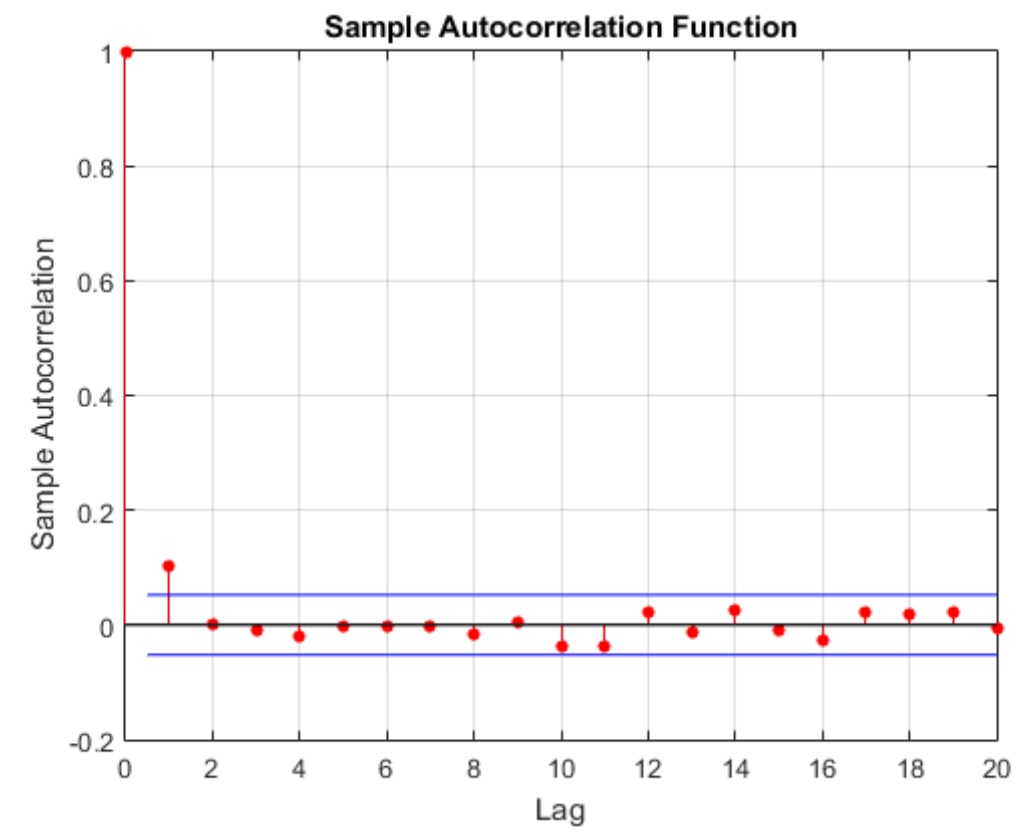}
		\caption{ACF coefficients for Period 3 (07/09 - 05/15) for BSE}
		\label{fig:combined}
	\end{minipage}
	\hspace{\fill}
	\begin{minipage}[t]{0.45\textwidth}
		\includegraphics[width=\linewidth]{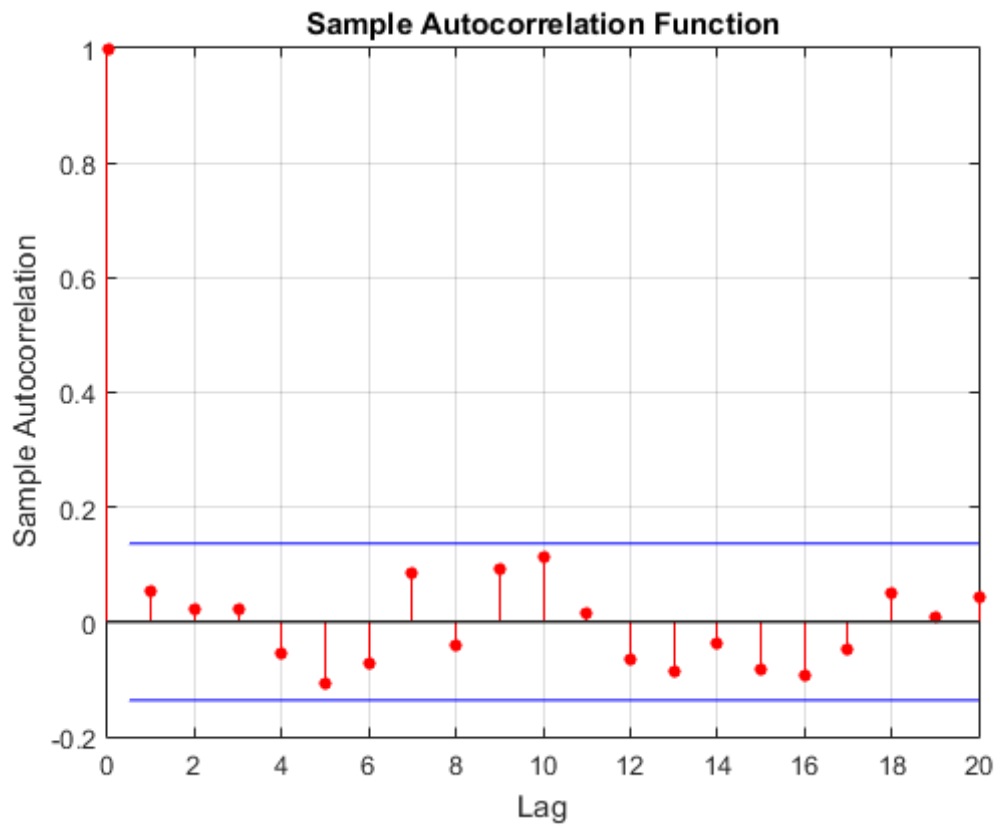}
		\caption{ACF coefficients for Period 4 (06/15 - 04/16) for BSE}
		\label{fig:combined}
	\end{minipage}
\end{figure}
\FloatBarrier

\subsection{Correlations between the market returns}

\FloatBarrier
\begin{table}[!htbp]
	\begin{adjustwidth}{0.0cm}{}
		\resizebox{\columnwidth}{!}{%
			\begin{tabular}{c|c|c|c|c|c} \hline
				& Daily & Period 1 (01/96-11/07) & Period 2 (12/07-06/09) & Period 3 (07/09-05/15) & Period 4 (06/15-04/16)\\ \hline
				Returns & -0.1318 &	0.1318 &	0.0403 & 0.3077 &	0.2315\\ \hline
				Prices & 0.7178 & 	0.6902 & 0.9118 & 	0.2679 & 	0.6359\\ \hline
			\end{tabular}%
		}%
		\caption{Correlations between the market prices and returns}\label{tbl:names}
	\end{adjustwidth}
\end{table}
\FloatBarrier
\noindent
As both the markets did not have equal no of working days in our period of analysis, the index closing price prevailing the last working day is considered to be closing price for that day for the market given the other market is open but the market in consideration is closed. And hence, the no of observations were equalised. We don’t find significant correlation between the daily returns from the indexes, although it shoots up during recession and then slowly starts decreasing. Whereas the daily closing prices show a very high correlation and provide a very interesting result. These seem to go up during the crisis periods and had almost become perfectly correlated during the global financial crisis but saw a drastic fall in the period after that. This can be explained due to increasing integration of the world market or the increase in trade in these periods. Also, this high correlation might partly explain the reason why these economies weren’t drastically hit during recession. 
\\
\\
And to find out more about this behaviour, we extract the country wise data for exports and imports between India and China in from the official website of Ministry of Commerce and Industry (Govt. of India, http://commerce.nic.in/). The total trade between the two countries  is calculated by summing up the exports and imports for the period 1997- 2015 and the graph for the same is plotted. From the trends seen, we can infer a direct relationship between the imports by India and the total trade and they have seen a rapid growth after 2005 and post recession period. The exports seem to be falling after 2011. The drop in trade in 2015 is because of the partly available data from the financial year 2015-2016 (Apr-Jan). Hence the trade data might help us a bit in explaining the correlation between the returns but fails to explain the extraordinary correlation between the closing prices.

\FloatBarrier
\begin{table}[!htbp]
	\begin{adjustwidth}{0.0cm}{}
		\resizebox{\columnwidth}{!}{%
			\begin{tabular}{c|c|c|c} \hline
				Year &	Exports (millions USD) &	Imports (millions USD) & Total (millions USD)\\ \hline
				2015 &	7516.1	&51811.23&	59327.33\\ \hline
				2014 &	11934.25&	60413.17&	72347.42\\ \hline
				2013 &	14824.36&	51034.62&	65858.98\\ \hline
				2012 &	13534.88&	52248.33&	65783.21\\ \hline
				2011 &	18076.55&	55313.58&	73390.13\\ \hline
				2010 &	14168.86&	43479.76&	57648.62\\ \hline
				2009 &	11617.88&	30824.02&	42441.9\\ \hline
				2008 &	9353.5	&32497.02&	41850.52\\ \hline
				2007 &	10871.34&	27146.41&	38017.75\\ \hline
				2006 &	8321.86	&17475.03&	25796.89\\ \hline
				2005 &	6759.1	&10868.05&	17627.15\\ \hline
				2004 &	5615.88	&7097.98&	12713.86\\ \hline
				2003 &	2955.08	&4053.21&	7008.29\\ \hline
				2002 &	1975.48	&2792.04&	4767.52\\ \hline
				2001 &	951.95	&2036.39&	2988.34\\ \hline
				2000 &	831.3	&1502.2&	2333.5\\ \hline
				1999 &	539.04	&1282.89&	1821.93\\ \hline
				1998 &	427.16 &	1096.71&	1523.87\\ \hline
				1997 &	717.95 &	1112.05&	1830\\ \hline
				1996 &	614.8 & 	756.91&	1371.71\\ \hline
			\end{tabular}%
		}%
		\caption{Imports and Exports between India and China during 1996 - 2015}\label{tbl:names}
	\end{adjustwidth}
\end{table}
\FloatBarrier

\begin{figure}[H]
	\hspace*{-2.5cm}   
	\centering
	\includegraphics[width=1.15\textwidth]{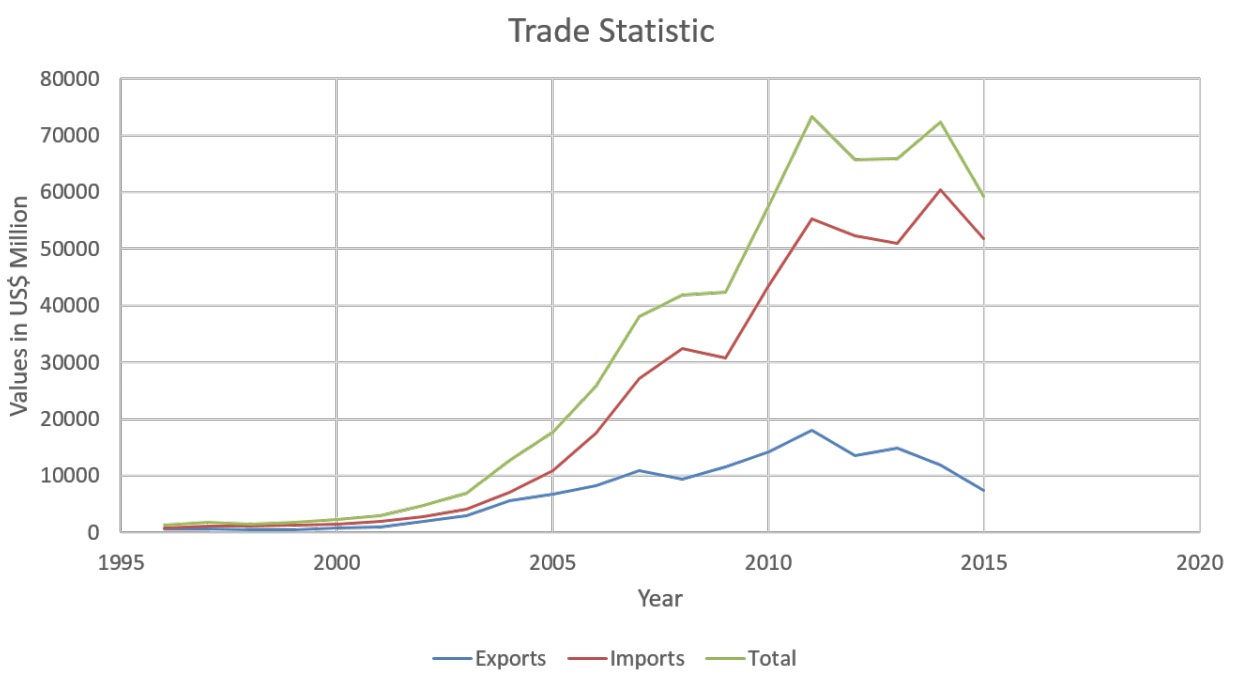}
	\caption{\label{fig:frog} Trade Statistics between India and China during 1996 - 2015}
\end{figure}

\subsection{Structural Changes}

In this paper, we have used The Hodrick–Prescott filter (HP Filter) to study the structural changes that arise in the time series consisting of the daily returns of both the BSE 200 and SSE Composite indices.

\begin{figure}[H]
	\hspace*{-3.1cm}   
	\centering
	\includegraphics[width=1.0\textwidth]{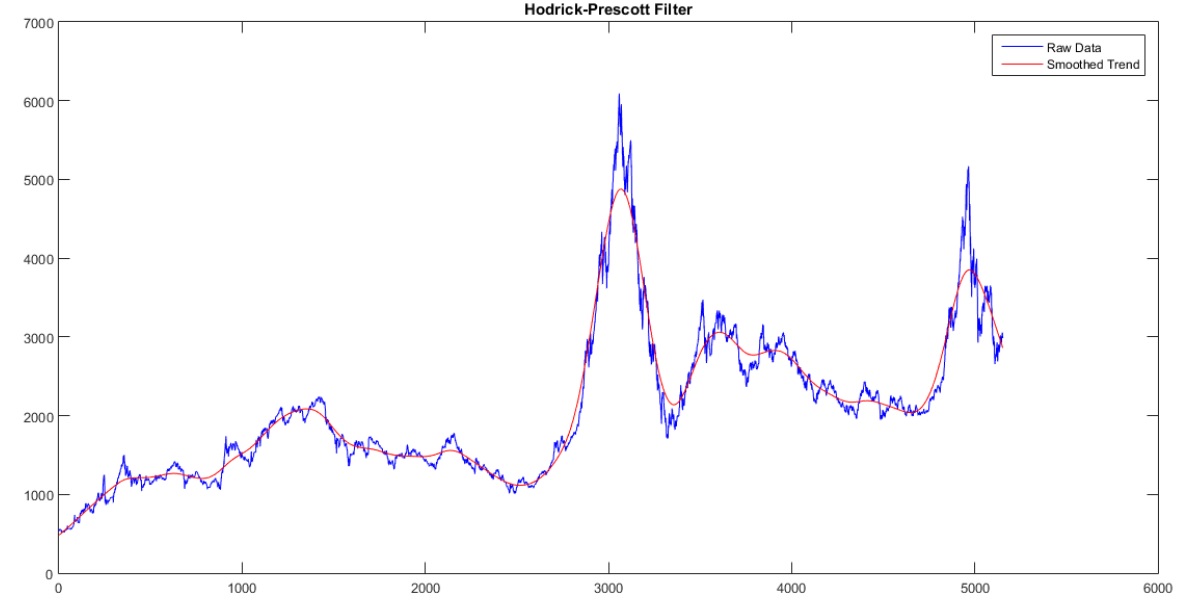}
	\caption{\label{fig:frog} Smoothing of series of returns of SSE Composite Index from 01/96 - 04/16}
\end{figure}

\begin{figure}[H]
	\hspace*{-3.1cm}   
	\centering
	\includegraphics[width=1.0\textwidth]{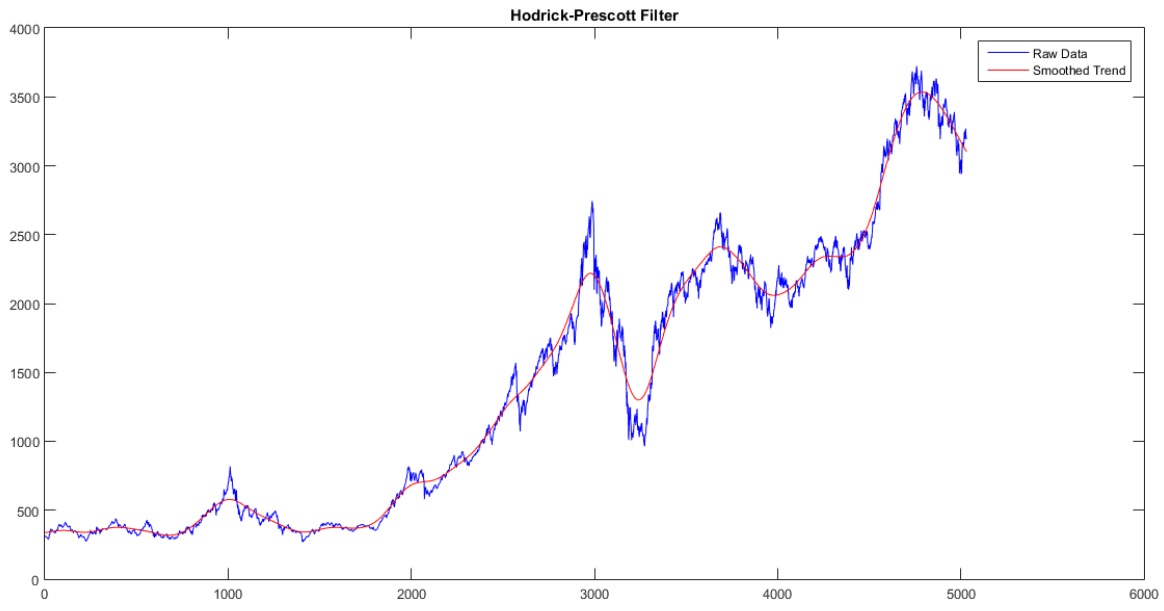}
	\caption{\label{fig:frog} Smoothing of series of returns of BSE Composite Index from 01/96 - 04/16}
\end{figure}
\noindent
Let $y_{t}$ be a time series. $y_{t}$ can be decomposed into two parts as follows:-
\begin{equation}
y_{t} = \tau_{t} + c_{t}
\end{equation}
where, $\tau_{t}$ is the trend component and $c_{t}$ is the cyclical component. 
The Hodrick–Prescott (HP) is a high-pass filter that is  used to separate a time-series $y_t$ into trend and cyclical components. The stationary behavior of the trend component is not guaranteed. It may be non-stationary and contain a stochastic trend. Using this type of filter one tries to obtain $c_{t}$ i.e. the cyclical component of the time series. This cyclical component is stationary and is driven by stochastic cycles at a range of periods. Figure 34 and 35 show the time series and its smoothed trend. The time series has been smoothed whnever there is an abrupt rise or fall in it. These periods correspond to periods of intense activity in the stock market where the stock market has either risen sharply or plunged severely. 
In the HP filter technique, one constructs $\tau{t}$ in such a way that the following quantity is minimized:-

\begin{equation}
\sum_{i=1}^{T}(y_{t} - \tau_{t})^2 + \lambda \sum_{i=2}^{T-1}[(\tau_{t+1} - \tau_{t}) - (\tau_{t} - \tau_{t-1})]^2
\end{equation}
The first term is the sum of the squared deviations of $y_{t}$ from the trend and the second term, which is the sum of squared second differences in the trend, is a penalty for changes in the trend’s growth rate. The larger the value of the positive parameter $\lambda$, the greater the penalty and the smoother the resulting trend will be. 
As a rule of thumb, $\lambda$ the smoothing parameter is calculated as follows:-
\begin{equation}
\lambda = 100*(PV)^2
\end{equation}
where, PV is the period value.
\\
Period values are as follows:-
\begin{itemize}
	\item Daily data - 365
	\item Monthly data - 12
	\item Quarterly data - 4
	\item Annual data - 1
\end{itemize}

\section{Limitations}
New set of findings may be worked out by taking data for different time periods since the study is longitudinal in nature. The results might also show variations if some other index is taken as the representatives for that market. Other additions to the study might include checking efficiencies through weekly and monthly analysis.  There have been many other instances of financial shocks globally in this time period which weren’t accounted for. 
\\
\\
The first period of our analysis is relatively bigger than the other periods and might lead to  few distortions in the results. Also, the analysis from the final period, i.e. period IV might not be completely reliable because of less number of data points(<200) and a clear conclusion could be drawn of the its entire effects only in the coming years when the market comes out of crisis.
\\
\\
This paper tries to study the impact of two recent financial crisis on the Indian and Chinese stock markets by taking the data for last twenty years on a daily basis and dividing it into four time periods according to the occurrence  of the financial crises. The present paper adds to the past literature of market efficiency by studying the impact of the financial crisis of 2008 and the recent Chinese crisis of 2015 on stock market efficiency in the leading emerging markets of India and China.
\\
\\
The literature on market efficiency is becoming extremely extensive that a careful analysis of it is beyond the scope of this thesis. Consequently, we provide a thorough but brief discussion of central findings in the market efficiency regarding random walk hypothesis or weak-form efficiency providing a general picture of this study.

\section{Conclusion}
The efficiency in stock markets explains the extent to which the stock prices reflects all available information in the market and therefore one can carry out fundamental analysis by relying upon this information to chart out a trading strategy for guaranteed returns. Our analysis from various tests help us in concluding that both the Indian and Chinese stock markets do not exhibit weak form market efficiency and thus do not follow random walk overall. One common observation was inefficiency of these market before the recession hit the world economy. However, the shocks in the form of crises are helping the market move towards efficiency slowly. The recent Chinese financial crisis did not impact the behavior of Chinese stock markets to a great extent but did have an impact on the Indian counterpart. Similarly, the global financial crises helped the Chinese market move towards efficiency but their recent crisis has countered that growth. Indian markets seem be the better of the two with slow but gradual shift towards following random walk (ACF test) and being less volatile than the Chinese counterpart. But these results only point out that there are possibilities of earning extra income in both the markets because abnormal returns are possible only when the market is inefficient as the future prices can be predicted using the past information. 

\begin{appendices}

	\section{Runs Test}
	The runs test determines whether successive price changes are independent and unlike the serial correlation test of independence, is non-parametric and does not require returns to be normally distributed. Observing the number of ‘runs’ - or the sequence of successive price changes with the same sign - in a sequence of price changes tests the null hypothesis of randomness. In the approach selected, each return is classified according to its position with respect to the mean return.
	\\
	\\
	That is, a positive change is when the return is greater than the mean, a negative change when the return is less than the mean, and zero change when the return equals the mean. To perform this test, A is assigned to each return that equals or exceeds the mean value and B for the items that are below the mean. Let $n_A$ and $n_B$ be the sample sizes of items A and B respectively. The test statistic is U -  the total number of runs. For large sample sizes, that is where both $n_A$ and $n_B$ are greater than twenty, the test statistic is approximately normally
	distributed (Berenson and Levine 2002):
	
	\begin{equation}
	Z = \frac{U - \mu_{U}}{\sigma_{U}}
	\end{equation}
	
	where, $\mu_{U}$ = $\frac{2n_{A}n_{B}}{n} + 1$, $\sigma_{U}$ = $\sqrt{\frac{2n_{A}n_{B}(2n_{A}n_{B}-n)}{n^2(n-1)}}$

	\section{ADF Test}
	The subsequent research about the market efficiency has used a new methodology to test the random walk nature of stock prices that is known by unit root test. This methodology was developed by Dickey and Fuller (1981) is used to examine the stationarity of the time series. The unit root test is designed to discover whether the series is difference-stationary (the null hypothesis) or trend-stationary (the alternative hypothesis) (Campbell et al.
	1997, 65). 
	\\
	\\
	A series with unit root is said to be non-stationary indicating nonrandom
	walk. The most commonly used test to examine the existence of a
	unit root is the Dickey-Fuller test. This unit root test provides evidence on
	whether the stock prices in Chinese stock market follow a random walk.
	Therefore, it is also a test of the weak-form market efficiency.
	The standard Dickey-Fuller (DF) test is appropriate for a series generated by
	an autoregressive process of order one, AR(1).
	\\
	\\
	If, however, the series follows an AR(p) process where p>1, the error term in the standard DF test will be autocorrelated. Autocorrelated will invalidate the use of the DF test distribution, which is based on the assumption that the error term is white noise. The Augmented Dickey-Fuller (ADF) test includes additional lagged
	difference terms to account for this problem. (Eviews 2004; Dickey and Fuller
	1981) The ADF unit root test is based on the following regression:
	
	\begin{equation}
	\Delta P_{t} = \gamma P_{t-1} + \sum_{i=1}^{q}(\rho_{i}\Delta P_{t-i}) + \epsilon_{t}
	\end{equation}
	
	\begin{equation}
	\Delta P_{t} = \mu + \gamma P_{t-1} + \sum_{i=1}^{q}(\rho_{i}\Delta P_{t-i}) + \epsilon_{t}
	\end{equation}
	
	\begin{equation}
	\Delta P_{t} = \mu + \alpha_{1}t \gamma P_{t-1} + \sum_{i=1}^{q}(\rho_{i}\Delta P_{t-i}) + \epsilon_{t}
	\end{equation}
	where $\delta$ represents first differences and $P_{t}$ is the log of the price index, $\mu$ is the constant, $\gamma$ and $\rho$ are coefficients to be estimated, q is the number of lagged terms, t is the trend, 1 a is the estimated coefficient for the trend, and the error term t, $\epsilon$ is assumed to be white noise. Then the null hypothesis to be tested is:
	$H_0$: $\rho$ = 0 (Non-stationary or unit root)
	$H_1$: $\rho$ < 0 (Stationary or no unit root)
	The first equation (3) is a pure random walk model without constant and time trend, the second equation (4) is with constant and without time trend, and third eqaution (5) includes
	both the constant and time trend. 
	\\
	\\Accordingly, equation (13b) tests for the null hypothesis of a random walk against a stationary alternative, while equation (13c) tests for the same null against a trend stationary alternative. To test the significance of the estimated $\gamma$ coefficients, the Augmented Dickey-Fuller unit root test is computed the tau statistic ( $\tau$ ) for each estimated coefficient, in the same way as a student’s t-statistic is calculated.
	\\
	\\
	However, the estimated $\tau$ values do not follow the same distribution as
	student’s t. The statistical significance of the estimated $\tau$ values must be
	assessed by comparing them with critical values derived for the $\tau$ distribution tabulated in Dickey and Fuller (1981). If the estimated $\tau$  value is less than the critical value in absolute terms, then the null hypothesis of the existence of unit root cannot be rejected.
	\\
	\\
	It is important to notice that the appropriate critical values depend on sample size, since as in most of hypothesis tests, for any given level of significance the critical values of the t-statistic decrease as sample size increases. (Enders 2004, 182; Eviews 2004) MacKinnon (1991) estimates the calculation of Dickey-Fuller critical values for any sample sizes is hereby used here (Eviews 2004).

	\section{Autocorrelation test (ACF)}
	Autocorrelation estimates may be used to test the hypothesis that the process
	generating the observed return is a series of i.i.d random variables. It helps to evaluate whether successive values of serial correlation are significantly different from zero. To test the joint hypothesis that all autocorrelation coefficients $q_k$ are simultaneously equal to zero, Ljung and Box’s (1978) portmanteau Q-statistic is used in the study. The test statistic is defined as:
	
	\begin{equation}
	ACF(k) = \frac{\sum_{t=1-k}^{n}(y_{t}-\bar{y})(y_{t-k}-\bar{y})}{\sum_{t=1}^{n}(y_{t}-\bar{y})^2}
	\end{equation}
	where $\bar{y}$ is the overall mean of the concerned series with ‘$n$’ observations.
	
	The standard error of $ACF($k$)$ is given by:-
	\begin{equation}
	se_{ACF(k)} = \frac{1}{\sqrt{n-k}}
	\end{equation}

	If $n$ is sufficiently large ($n$ $≥$50), the standard error of $ACF($k$)$ can be approximated to:-
	\begin{equation}
	se_{ACF(k)} = \frac{1}{\sqrt{n}}
	\end{equation}
	The following t-distribution will be used to test the hypothesis whether$ACF($k$)$ is significantly different from zero or not.
	\begin{equation}
	t = \frac{ACF(k)}{se_{ACF(k)}}
	\end{equation}
	
\end{appendices}

\end{document}